\documentclass[iop]{emulateapj} 

\usepackage{verbatim}

\newcommand{\msun}{$M_{\odot}$}
\newcommand{\lsun}{$L_{\odot}$}

\newcommand{\rxj}{RX~J0152-13} 
\newcommand{\sfr}{$M_{\odot}$~yr$^{-1}$}

\newcommand{\galmpc}{Mpc$^{-2}$}
\newcommand{\pmpc}{Mpc$^{-1}$}
\newcommand{\kmps}{km~s$^{-1}$}

\newcommand{\uJy}{$\mu$Jy}

\newcommand{\oii}{[OII]$\lambda3727$~\AA}
\newcommand{\av}{A$_{\rm V}$}

\newcommand{\filters}{$BVRizK_s$}
\newcommand{\zspec}{$z_{\rm spec}$}
\newcommand{\zldp}{$z_{\rm LDP}$}
\newcommand{\zwindow}{$0.80<z<0.87$}
\newcommand{\zwindowwide}{$0.6<z<0.9$}
\newcommand{\mcut}{10.25}
\newcommand{\ntot}{3326}   
\newcommand{\nmcutzmag}{1174} 
\newcommand{\nmcut}{977} 
\newcommand{\nldp}{9341} 
\newcommand{\nmipsks}{638} 
\newcommand{\mcutrange}{$M>1.8 \times 10^{10}$~\msun}
\newcommand{\mcutrangelog}{$\log M/M_{\odot} >~$\mcut}

\begin{document}

\title{The Star Formation Rate-Density Relation at $0.6 < \lowercase{z} < 0.9$ and the Role of Star Forming Galaxies \altaffilmark{1,2,3,4,5}}

\author{Shannon G. Patel \altaffilmark{6,7},
Daniel D. Kelson\altaffilmark{8},
Bradford P. Holden\altaffilmark{6},
Marijn Franx\altaffilmark{7},
Garth D. Illingworth\altaffilmark{6}
}

\altaffiltext{1}{This paper includes data gathered with the 6.5~meter Magellan Telescopes located at Las Campanas Observatory, Chile.}
\altaffiltext{2}{This work is based on observations made with the Spitzer Space Telescope, which is operated by the Jet Propulsion Laboratory, California Institute of Technology under a contract with NASA. Support for this work was provided by NASA through an award issued by JPL/Caltech.}
\altaffiltext{3}{Based in part on data collected at Subaru Telescope and obtained from the SMOKA, which is operated by the Astronomy Data Center, National Astronomical Observatory of Japan.}
\altaffiltext{4}{Some of the data presented herein were obtained at the W. M. Keck Observatory, which is operated as a scientific partnership among the California Institute of Technology, the University of California and the National Aeronautics and Space Administration. The Observatory was made possible by the generous financial support of the W.M. Keck Foundation.}
\altaffiltext{5}{Based on observations made with the NASA/ESA Hubble Space Telescope, obtained at the Space Telescope Science Institute. STScI is operated by the Association of Universities for Research in Astronomy, Inc. under NASA contract NAS 5-26555.}
\altaffiltext{6}{UCO/Lick Observatory, University of California, Santa Cruz, CA 95064; patel@ucolick.org}
\altaffiltext{7}{Leiden Observatory, Leiden University, P.O. Box 9513, NL-2300 AA Leiden, Netherlands}
\altaffiltext{8}{Observatories of the Carnegie Institution of Washington, Pasadena, CA 91101}

\begin{abstract}
We study the star formation rates (SFRs) of galaxies as a function of local galaxy density at \zwindowwide.  We used a low-dispersion prism in IMACS on the 6.5~m Baade (Magellan I) telescope to obtain spectra and measured redshifts to a precision of $\sigma_z/(1+z) \sim 1\%$ for galaxies with $z_{\rm AB} <  23.3$~mag.  We utilized a stellar mass-limited sample of \nmcut\ galaxies above \mcutrange\ (\mcutrangelog) to conduct our main analysis.  With three different SFR indicators, (1) {\em Spitzer} MIPS 24~\micron\ imaging, (2) SED fitting, and (3) \oii\ emission, we find the median specific SFR (SSFR) and SFR to decline from the low-density field to the cores of groups and a rich cluster.  For the SED and [OII] based SFRs, the decline in SSFR is roughly an order of magnitude while for the MIPS based SFRs, the decline is a factor of $\sim 4$.  We find approximately the same magnitude of decline in SSFR even after removing the sample of galaxies near the cluster.  Galaxies in groups and a cluster at these redshifts therefore have lower star formation (SF) activity than galaxies in the field, as is the case at $z \sim 0$.  We investigated whether the decline in SFR with increasing density is caused by a change in the proportion of quiescent and star forming galaxies (SFGs) or by a decline in the SFRs of SFGs.  Using the rest-frame $U-V$ and $V-J$ colors to distinguish quiescent galaxies from SFGs (including both unattenuated blue galaxies and reddened ones) we find the fraction of quiescent galaxies increases from $\sim 32\%$ to $79\%$ from low to high density.  In addition, we find the SSFRs of SFGs, selected based on $U-V$ and $V-J$ colors, to decline with increasing density by factors of $\sim 5-6$ for the SED and [OII] based SFRs.  The MIPS based SSFRs for SFGs decline with a shallower slope.  The declining SFRs of SFGs with density is paralleled by a decline in the median \av, providing indirect evidence that the cold gas content that fuels future SF is diminished in higher density environments.  The order of magnitude decline in the SSFR-density relation at \zwindowwide\ is therefore driven by both a combination of declining SFRs of SFGs as well as a changing mix of SFGs and quiescent galaxies.
\end{abstract}


\section{Introduction}

Hierarchical structure formation results in more massive dark matter halos assembling at late times.  Simulations predict that halos that host $L^{\star}$ galaxies, groups, and clusters, grow in mass by factors of $\sim 2-3$ between $z \sim 1$ and $z \sim 0$ \citep[e.g.,][]{wechsler2002,vandenbosch2002}.  At $z \sim 0$, these massive halos, or high density environments, typically host more evolved galaxy populations.  The morphological study of \citet{dressler1980} provided the first quantitative glimpse of this fact, finding elliptical and S0, or early type galaxies, to dominate in high-density regions of clusters while spirals and irregulars, or late-types, dominate at lower densities.  This morphology density relation (MDR) has also been found to extend to group environments at $z \sim 0$ \citep{postman1984}.  These high density environments have continuously accreted galaxies from a range of lower density environments over time.  Thus, mass growth in such high density environments would lead galaxies within newly accreted halos at $z \sim 1$ to undergo various transformations in their properties if the general environmental trends at $z \sim 0$ hold out to these higher redshifts.

In addition to morphology other galaxy properties such as rest-frame colors \citep[e.g.,][]{kauffmann2004,hogg2004,blanton2005b,baldry2006} and star formation rates \citep[SFRs, e.g.,][]{abraham1996,hashimoto1998,gomez2003,kauffmann2004} have also been found to correlate with the local galaxy density at $z \sim 0$.  Large redshift surveys find such correlations in the field as well, indicating the importance of mechanisms that operate at halo mass scales below that of clusters in transforming galaxy properties at $z \sim 0$.  In general the works above find red, quiescent, early-type galaxies represent a larger proportion of the population in higher density regions at $z \sim 0$ such as the cores of groups and clusters, while blue, star-forming, late-type galaxies make up a larger share in lower density regions.

While several studies at $z \sim 0$ have found the mean or median SFRs of galaxies to be lower at higher densities such as in groups and clusters \citep[e.g.,][]{hashimoto1998,gomez2003,kauffmann2004}, at higher redshifts the direction of this SFR-density trend has been the subject of debate.  For luminosity-limited samples, \citet{elbaz2007} and \citet{cooper2008} find that at $z \sim 1$ the SFR-density relation reverses, such that galaxies at higher densities have higher SFRs.  Meanwhile, for a stellar mass-limited sample at $z \sim 0.8$, \citet{patel2009b} found the obscured SFR-density relation to decline at higher densities, much like at $z \sim 0$.  The direction of the SFR-density trend at $z \sim 1$ has implications for the effectiveness of the various physical processes that operate in different environments in regulating star formation (SF) and evolving galaxies.  For example, a reversed SFR-density trend at $z \sim 1$ would imply that the physical processes that lead to the shutdown of SF in high density group environments produce fewer red, non-star forming galaxies relative to a declining SFR-density relation.

Studying the origin of the SFR-density relation provides additional insight into the star formation histories (SFHs) of galaxies across a range of environments.  For example, a declining SFR-density relation at a fixed stellar mass can be produced by either (1) the proportion of quiescent and star forming galaxies (SFGs) changing with density, (2) the SFRs of SFGs decreasing at higher densities, or (3) some combination of these two scenarios.  The fraction of red and/or blue galaxies are often used to gauge scenario (1) above \citep[e.g.,][]{baldry2006,cucciati2006,gerke2007,cooper2007,patel2009}, however, dusty SFGs can display red optical colors \citep{vanderwel2007b,maller2009}, thus complicating such measurements.  With extinction in galaxies also dependent on environment \citep{kauffmann2004}, computing the fraction of quiescent and SFGs vs. local density from a single optical color is further complicated.  Thus, selecting SFGs, both the unobscured and obscured variety, is essential in determining the contribution of the three scenarios listed above in producing the SFR-density relation.

In this work, we build on our analysis in \citet{patel2009b} in which we found declining obscured SSFR-density and SFR-density relations at a fixed stellar mass at $z \sim 0.83$.  Here, we expand the analysis with three different SFR indicators in order to confirm these declining relations found in \citet{patel2009b}.  Our sample in this work also spans a larger redshift range (\zwindowwide) than in \citet[][\zwindow]{patel2009b} and encompasses galaxies in several groups and a cluster.  As a consequence, we are able to determine how the SFRs of galaxies vary across a range of larger dark matter halos.  Finally, one of our primary goals in this work is to determine how the SFRs of individual galaxies change in order to produce the declining SFR-density relation.  To this end, we employ a color-color diagram to distinguish quiescent and SFGs, as also utilized by several other recent works \citep[e.g.,][]{wuyts2007,williams2009}.

In conducting this study, we primarily utilize stellar mass-limited samples as opposed to luminosity-limited samples.  A limiting magnitude for a survey implies a limiting mass for old, quiescent galaxies that is generally higher than the limiting mass for young, SFGs at a given redshift (see, e.g, Appendix~\ref{app_masslimit}).  As a consequence, below the stellar mass limit implied by passive evolution of old systems, no sample can be used to definitively characterize the average galaxy.  This does not mean that the lower mass, blue SFGs found in luminosity-limited samples cannot be used to study particular modes of star formation (SF), but merely that such samples are not necessarily representative of the full population at low mass.  These different selections can therefore lead to widely varying results.  For example, in studying the mass-limited morphology-density relation (MDR) in clusters, \citet{holden2007} found little evolution in the fraction of early-type galaxies at a fixed density between $z \sim 1$ and $z \sim 0$ \citep[see also,][]{holden2006}.  An extension of this analysis to field densities also results in a similar conclusion \citep{vanderwel2007b}.  In contrast, previous results that utilized luminosity-limited samples found a significant buildup of early-type galaxies at late times, with the share of S0s growing at the expense of spirals \citep[e.g.,][]{dressler1997,postman2005}.  Other galaxy properties correlate with stellar mass as well, such as optical colors \citep{bell2001,kauffmann2003}, SFRs \citep{brinchmann2004,salim2007,noeske2007}, and extinction \citep{garn2010b}.  In drawing conclusions, it is therefore important to consider the distribution of stellar masses for different sub-samples that are being compared.  

Our $z$-band spectroscopic selection ($z_{\rm AB}<23.3$~mag) allows us to build mass-limited samples at $z \sim 0.9$ down to $M>0.2 M^{\star}$ \citep[$M^{\star} \approx 10^{11}$~\msun,][]{fontana2006}, probing an important stellar mass range that is responsible for much of the buildup on the red-sequence since $z \sim 1$ \citep{brown2007}.  This relatively low stellar mass limit is a key advantage of our survey over other large spectroscopic field and cluster surveys at these redshifts.  For comparison, the stellar mass limit at $z \sim 0.9$ for the $R_{\rm AB}<24.1$~mag selected DEEP2 survey is $M \approx 5 \times 10^{10}$~\msun\ \citep[see, e.g.,][]{noeske2007}.  For the $I_{\rm AB}<22.5$~mag selected zCOSMOS-bright survey the stellar mass limit is $M \approx 7 \times 10^{10}$~\msun\ \citep[see, e.g.,][]{lilly2009}.  For the $I<23$~mag (Vega) selected EDisCS survey \citep{white2005} the stellar mass limit is $M \approx 3 \times 10^{10}$~\msun.  The $I_{\rm AB}<24$~mag selected VVDS-Deep survey \citep{lefevre2004} reaches a comparable mass limit at $z \sim 0.9$ to that in this work.

Our paper is outlined as follow.  In \S~\ref{sfr_sec_data} we discuss the data used while in \S~\ref{sfr_sec_analysis} we derive key quantities such as stellar masses, SFRs, and rest-frame colors.  We present our method for distinguishing quiescent and SFGs in \S~\ref{sfr_sec_bicolor}.  Results are presented in \S~\ref{sfr_sec_sfrdensity} along with a discussion in \S~\ref{sfr_sec_discussion}.  Finally, a summary is presented in \S~\ref{sfr_sec_summary}.

We assume a cosmology with $H_0=70$~\kmps~\pmpc, $\Omega_M = 0.30$, and $\Omega_{\Lambda} = 0.70$.  Stellar masses and SFRs are based on a Chabrier IMF \citep{chabrier2003}.  All magnitudes are reported in the AB system unless otherwise stated.

\bigskip

\section{Data} \label{sfr_sec_data}

\subsection{Imaging}

\subsubsection{Optical and Near-IR Imaging from the Ground}

The analysis in this paper is based on imaging from multiple telescopes and instruments that span from the near-UV to mid-IR.  We obtained archival Subaru Suprime-Cam \citep{miyazaki2002b} imaging of \rxj\ over a wide field of size $29\arcmin \times 39\arcmin$ \citep[for a detailed discussion of the Suprime-Cam data, see][]{kodama2005}.  The field was observed in $VRiz$ with Suprime-Cam and had a seeing FWHM$\sim 0\farcs65$ in all four bands.  Limiting magnitudes were determined by placing $D=3\arcsec$ apertures across each image and fitting a Gaussian to the distribution of fluxes for the subset of apertures that did not fall on detected sources.  The $1\sigma$ width of the Gaussian fit was taken to represent the RMS in the background within the aperture.  The resulting $5\sigma$ depths in $VRiz$ were 25.72, 25.65, 25.20, 24.24 AB mag respectively ($D=3\arcsec$ aperture diameter).  Note that the $z$-band imaging was used for the spectroscopic selection and is discussed below.

We obtained $K_s$ imaging of the central $26\arcmin \times 26\arcmin$ using the Wide Field Infrared Camera \citep[WIRC,][]{persson2002} on the $2.5$~m DuPont telescope at Las Campanas.  The effective seeing of the mosaic was $\sim 0\farcs7$ and the $5\sigma$ depth was $21.22$~AB mag ($D=3\arcsec$ aperture).  The $K_s$-band imaging is important for constraining stellar mass estimates, as it samples the rest-frame $J$-band for most galaxies in our sample.  When combined with the optical imaging, the long wavelength baseline afforded by the $K_s$ data also enables one to constrain the amount of dust extinction. 

The Magellan IMACS \citep{dressler2006} $B$-band imaging covers the full Suprime-Cam FOV.  The seeing was $1\farcs2$ and the $5\sigma$ depth was 24.55~AB mag ($D=3\arcsec$ aperture).  The $B$-band imaging probes the rest-frame UV continuum and therefore provides a measure of the unobscured SFR.  

The $BVRizK_s$ imaging are binned to the same pixel scale ($0\farcs2$), and placed on the same astrometric solution.  The $z$-band image was used for source detection, and the ``double-image mode'' of SExtractor \citep{bertin1996} was used to extract fluxes in matched apertures in $BVRizK_s$.

\subsubsection{Space Based Mid-IR and Optical Imaging}

We obtained {\em Spitzer} MIPS \citep{rieke2004} 24~\micron\ imaging of the field in Cycles 2 and 5, and combined the data with existing MIPS imaging of the central regions of the field \citep[see][]{marcillac2007}.  The MIPS coverage in this paper with our Cycle 2 and 5 data extends well beyond the region centered on the core of \rxj.  This data was also employed in \citet{patel2009b}.  The MIPS imaging covers a total of $\sim 0.1$~deg$^2$.  We used MOPEX \citep{makovoz2005} to reduce the MIPS data and to create the final mosaic, which had a pixel scale of $1.245\arcsec$.  The PSF FWHM was $\sim 6\arcsec$.  The APEX module in MOPEX was used for source detection.  The $5\sigma$ detection limit was $125$~\uJy\ for a $D=6\arcsec$ aperture after applying a factor of $3.0$ aperture correction to a total flux for point sources.  MIPS sources were matched with the $z_{\rm AB}<23.3$~mag catalog using a $1\farcs5$ matching distance.  Our choice of matching distance reflects a trade off between the number of matched objects and contamination due to multiple matches.  The matching distance is roughly equivalent to two times the RMS scatter in the position offset between MIPS objects and objects in the $z$-band catalog.  Overall, $\sim 19\%$ of objects from the $z_{\rm AB}<23.3$~mag catalog with MIPS coverage are matched to a MIPS source.  In cases where multiple objects in the $z$-band catalog could be matched with the same MIPS source, the closest object was assigned as the match.  Of the objects assigned to a MIPS counterpart, $\sim 4\%$ of them were assigned in this way.  Note that not all galaxies are matched with a MIPS counterpart due to a combination of (1) galaxies with MIPS fluxes that are likely below the detection limit and (2) centroid shifts due to the confusion limit that lead to non-matches \citep[see, e.g.,][]{hogg2001}.  The MIPS data are used to derive obscured SFRs as discussed in \S~\ref{sfr_sec_sfr}.

We also used HST ACS \citep{ford2003} imaging that was centered on \rxj.  The data reduction procedures are discussed in \citet{blakeslee2003}.  The imaging covers regions centered on the cluster core and outskirts ($R \lesssim 6\arcmin$). In this paper, we use F625W and F775W imaging of the central regions ($R \lesssim 3\arcmin$) and F606W and F814W imaging of the outskirts to make postage stamps (see \S~\ref{sfr_sec_bicolor}).

\subsection{Spectroscopy}
We selected galaxies with $z$-band MAG\_AUTO magnitudes of $z_{\rm AB} < 23.3$~mag for spectroscopy with IMACS on the Magellan/Baade telescope.  Note that this magnitude limit is much brighter than the $z$-band detection limit.  In place of the grating, we utilized a low dispersion prism (LDP) designed by S. Burles for use by the PRIMUS redshift survey \citep{coil2010}.  This configuration allowed us to place $\sim 3000$ slits, each with a width of $\sim 0\farcs8$, onto a single slitmask.  The wavelength range of the LDP spectra is $\sim 4500$~\AA\ to $\sim 1$~\micron, and spans roughly $\sim 100$~pixels.  The dispersion and resolution of the LDP vary strongly with wavelength.  At $5000$~\AA, the dispersion is $\sim 20$~\AA/pixel and the spectral resolution FWHM~$\sim 80$~\AA, while at $8500$~\AA, the dispersion is $\sim 150$~\AA/pixel and the spectral resolution FWHM~$\sim 600$~\AA.  The median total exposure time for each object was $\sim 3$~hr/pixel, roughly three times longer than PRIMUS.  The spectra were flux calibrated using spectrophotometric flux standards.  The data reduction and spectral extraction process for the LDP data are discussed in more detail in Patel~et~al. (2011, in preparation).

\subsection{Data Preparation for SED Fitting}

In preparation for fitting galaxy SEDs, which include both broadband photometry and prism spectroscopy, we took steps to account for differences between the imaging PSFs.  For each object, we used a $D=3\arcsec$ color aperture for the broadband photometry, which was aperture corrected to a total magnitude for all bands using the difference between the $z$-band MAG\_AUTO and $D=3\arcsec$ $z$-band magnitude.  While the $VRizK_s$ imaging had roughly similar seeing, the $B$-band seeing was slightly higher at FWHM$~\sim 1\farcs2$.  To account for the larger $B$-band PSF, we convolved the $V$ imaging to the $B$-band seeing of $1\farcs2$ and determined the difference in magnitude for the $D=3\arcsec$ aperture between the blurred and un-blurred $V$-band imaging.  This difference (typically $\sim 0.04$~mag) was added to the $D=3\arcsec$ $B$-band magnitude to ``correct'' it to the $VRizK_s$ seeing of $\sim 0\farcs7$.  

The LDP spectra were scaled to match the $VRi$ photometry.  The scaling at each LDP pixel was a combination of a constant value plus a wavelength dependent component, introduced to further improve upon the flux calibration computed from the spectrophotometric standard.

\section{Analysis} \label{sfr_sec_analysis}

\subsection{LDP Redshifts}

\begin{figure}
\epsscale{1.2}
\plotone{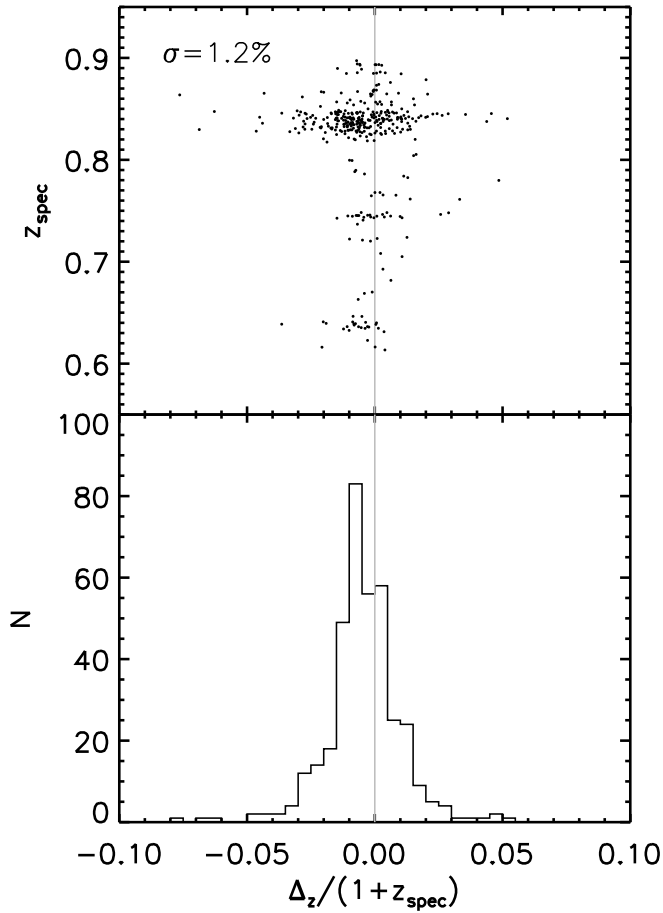}
\caption{Difference between LDP redshifts and redshifts from higher resolution spectroscopy ($\Delta_z = z_{\rm LDP} - z_{\rm spec}$), as a fraction of $(1+z_{\rm spec})$ versus \zspec\ (top).  Only the $377$ galaxies with $0.6<z_{\rm spec}<0.9$ and $z_{\rm AB} < 23.3$~mag are shown.  The bottom panel shows a histogram of $\Delta_z/(1+z_{\rm spec})$ for this sample.  The $1\sigma$ biweight scatter of $\Delta_z/(1+z_{\rm spec})$ is $\sigma = 1.2\%$.  The proportion of catastrophic outliers is very low, with only $1.3\%$ of the sample having $|\Delta_z|/(1+z_{\rm spec}) > 5\%$. \label{sfr_fig_redshifts}}
\end{figure}

LDP redshifts are determined by fitting the SEDs of galaxies with stellar population synthesis models.  Both the LDP spectra and broadband photometry are used in the fitting.  The method is briefly discussed in \citet{patel2009} and will be discussed in more detail in Patel~et~al. (2011, in preparation).

Our survey with the LDP finds galaxies spanning a range of redshifts.  The analysis in this work focuses on the redshift interval \zwindowwide.  The low redshift boundary was determined such that the bright magnitude selection criterion ($z_{\rm AB}>18$~mag) would not result in the loss of massive galaxies.  Meanwhile, the high redshift boundary was mostly determined by the desired stellar mass limit of our sample, with some consideration also for the larger redshift uncertainties at higher redshifts.

To determine the precision of the LDP redshifts, \zldp, we compare them to redshifts from higher resolution spectroscopy, \zspec, using the catalogs of \citet{tanaka2006} and \citet{demarco2005}.  We also include redshifts from our own Magellan/IMACS/LDSS3 and Keck/DEIMOS \citep{faber2003} spectroscopy in comparing \zldp\ to \zspec\ in Figure~\ref{sfr_fig_redshifts}.  For galaxies in our redshift range, $0.6<z_{\rm spec}<0.9$, the LDP redshifts have a biweight scatter of $\sigma_{z_{\rm LDP}-z_{\rm spec}}/(1+z_{\rm spec}) = 1.2\%$.  The scatter is approximately the same for blue and red galaxies.  Red galaxies have slightly lower \zldp\ compared to \zspec\ ($<1\%$ of $(1+z)$), accounting for the non-zero value of the mode in Figure~\ref{sfr_fig_redshifts}.

In reporting parameters from SED fitting in \S~\ref{sfr_sec_sed}, we used a different set of templates from those used for determining the redshift.  The templates used for determining redshifts utilized specific priors on various components (e.g. $M/L$, rest-frame colors, emission line ratios, etc.) in order to minimize the scatter in \zldp$~-~$\zspec.  Thus, these templates were used to optimize redshift measurements.  In addition, new data products ($BK_s$) have been incorporated into our analysis since the redshift determination process, and we have therefore employed more flexible SFHs in fitting these data (see Figure~\ref{sfr_fig_sedfits}).

\subsection{Spectroscopic Completeness}

Understanding the completeness of a spectroscopic survey is critical in reliably computing various quantities, such as local galaxy densities or the typical SFR or fraction of objects of a certain type in a given density bin.  We determine the spectroscopic completeness of our $z_{\rm AB}<23.3$~mag selected survey as a function of the $z$-band magnitude, $R-z$ color, and position on the sky (see Figure~\ref{fig_compmap} in the Appendix).  Overall we obtained a high quality LDP redshift for \nldp\ galaxies with $z_{\rm AB}<23.3$~mag.  At a given position on the sky, the spectroscopic completeness fraction in a magnitude-color bin is determined by taking the number of galaxies with redshift measurements, \zldp, and dividing by the number of galaxies in the same magnitude-color bin from the $z_{\rm AB}<23.3$~mag selection catalog.  The overall spectroscopic completeness of the survey is $\sim 74\%$.  The completeness is fairly uniform for bright and faint objects, varying from $\sim 76\%$ at $z_{\rm AB}<22.5$~mag to $\sim 71\%$ at $z_{\rm AB}>22.5$~mag.  Near the magnitude selection limit, $23 < z_{\rm AB} < 23.3$~mag, the completeness is slightly lower ($\sim 60-70\%$).  The completeness is also uniform in color at $\sim 74\%$ for $R-z>1$~AB mag and $R-z<1$~AB mag.  The completeness is slightly higher in the central regions ($R \lesssim 15\arcmin$: $\sim 81\%$) compared to the outer regions of the LDP spectroscopy ($R \gtrsim 15\arcmin$: $\sim 70\%$).  Where appropriate, measured quantities are assigned weights based on these completeness maps.

\bigskip
\bigskip

\subsection{SED Fitting} \label{sfr_sec_sed}

\begin{figure}
\epsscale{1.2}
\plotone{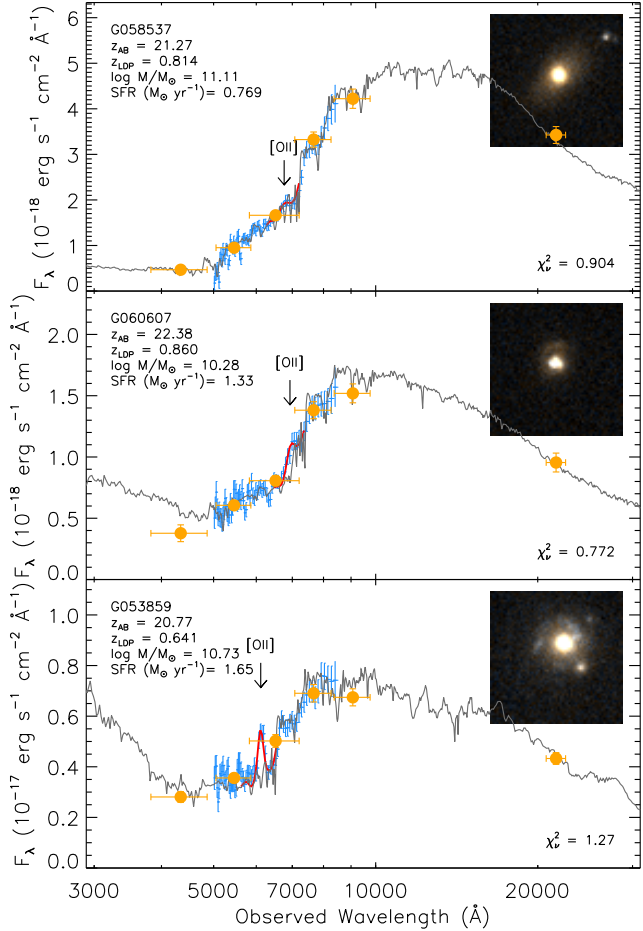}
\caption{Example SED fits to LDP spectroscopy (blue) and \filters\ broadband photometry (orange).  The best fitting redshifted full resolution BC03 $\tau$-model is shown in gray (NOTE: the actual data are fit to models that are convolved with the LDP resolution).  The Gaussian fit to \oii\ is indicated by the arrow and shown in red.  An HST ACS color postage stamp ($5\arcsec$ on a side) is shown inside each panel.  The LDP redshift, \zldp, and redshift from higher resolution spectroscopy, \zspec, are also indicated in each panel.  With the exception of the MIPS 24~\micron\ data, this figure showcases all of the data sets used in this paper.  The SF activity varies from quiescent on top to very active on bottom. \label{sfr_fig_sedfits}}
\end{figure}

In order to measure rest-frame properties of our sample, we fit their SEDs with \citet[hereafter BC03]{bc03} stellar population synthesis models (low resolution models), using a Chabrier IMF.  The $50$th percentile value of \zldp\ from the redshift likelihood function was used as the redshift of each galaxy and the subset of models from our grid (see below) with this redshift were used in the SED fitting.  Models were fit to both the LDP spectroscopy and \filters\ broadband photometry.  Due to the IMACS detector response and uncertainty in the flux calibration at redder wavelengths, we utilized the LDP wavelength range of $5000<\lambda<8500$~\AA\ in the SED fitting.  

We used a grid of $\tau$-models that span a wide range of galaxy SFHs.  Such $\tau$-models SFHs are commonly used in fitting SEDs \citep[e.g.,][]{papovich2001,forster2004b,franx2008}.  The grid consists of 4 parameters: (1) 3 metallicities: 0.4$Z_{\odot}$, $Z_{\odot}$, 2.5$Z_{\odot}$, (2) 10 values for $\tau$ logarithmically spaced between $0.1<\tau<20$~Gyr, (3) 20 ages logarithmically spaced between $0.3<t<7$~Gyr (for a given redshift, ages are limited to the subset with values less than the age of the universe at that epoch), and (4) 10 values for \av\ spaced between $0<$\av$<2$ and assuming a \citet{calzetti2000} extinction curve.  The stellar component is combined with a Gaussian representative of \oii\ in a non-negative least squares (NNLS) fit to the SED.  In carrying out the SED fitting, several important quantities are stored from the best-fitting minimum $\chi^2$ model, including the stellar mass, SFR, and [OII] flux from the Gaussian emission line component.  Figure~\ref{sfr_fig_sedfits} shows example fits to SEDs, which include LDP spectroscopy and $BVRizK_s$ broadband photometry.

\subsection{Rest-frame Colors and Magnitudes}

We use the procedure described in \citet{rudnick2003} to apply $K$-corrections and determine rest-frame colors and magnitudes.  The procedure is commonly used by other works to apply $K$-corrections \citep[see, e.g.,][]{taylor2009b,williams2009}.  To determine the magnitude of an object in a desired rest-frame filter, this method uses the redshift to find the two closest observed filters.  The observed color of the object from these two filters is used to search the grid of models from the SED fitting for the two models with the closest colors that straddle the observed color.  We interpolate in these two models, using the observed color, to determine the magnitude of the object in the desired rest-frame filter.  This procedure produces similar colors and magnitudes at $z \sim 0.8$ as in \citet{patel2009,patel2009b}, but has the advantage of allowing us to easily expand the analysis to suit the larger redshift range studied in this work.  The rest-frame filters relevant to this work are $UVJ$.  We use the Buser $UV$, and 2MASS $J$ filter curves supplied with BC03 in computing rest-frame magnitudes.

\subsection{Stellar Masses}

Stellar masses are one of the key parameters computed in the SED fitting.  We primarily use a stellar mass limited sample to conduct our study in this paper.  Given the $z$-band spectroscopic selection, in our redshift interval of interest, the limiting stellar mass is determined by the masses of the faint red galaxies.  Although the $z$-band magnitude limit results in blue galaxies with lower masses, their red counterparts of the same low mass do not make it into the sample.  We illustrate and discuss this further in the Appendix.  We find the limiting stellar mass at $z=0.9$ is \mcutrange\ (\mcutrangelog).  Of the \ntot\ galaxies in the redshift interval \zwindowwide\ with $z_{\rm AB}<23.3$~mag, \nmcutzmag\ are above the mass limit.  Of these, \nmcut\ have $K_s$-band imaging.  About $\sim 16\%$ of this mass-limited sample with $K_s$-band imaging is located in the vicinity of the cluster \rxj\ (\zwindow\ and $R<3$~Mpc).

\subsection{Local Density}

\begin{figure}
\epsscale{1.3}
\plotone{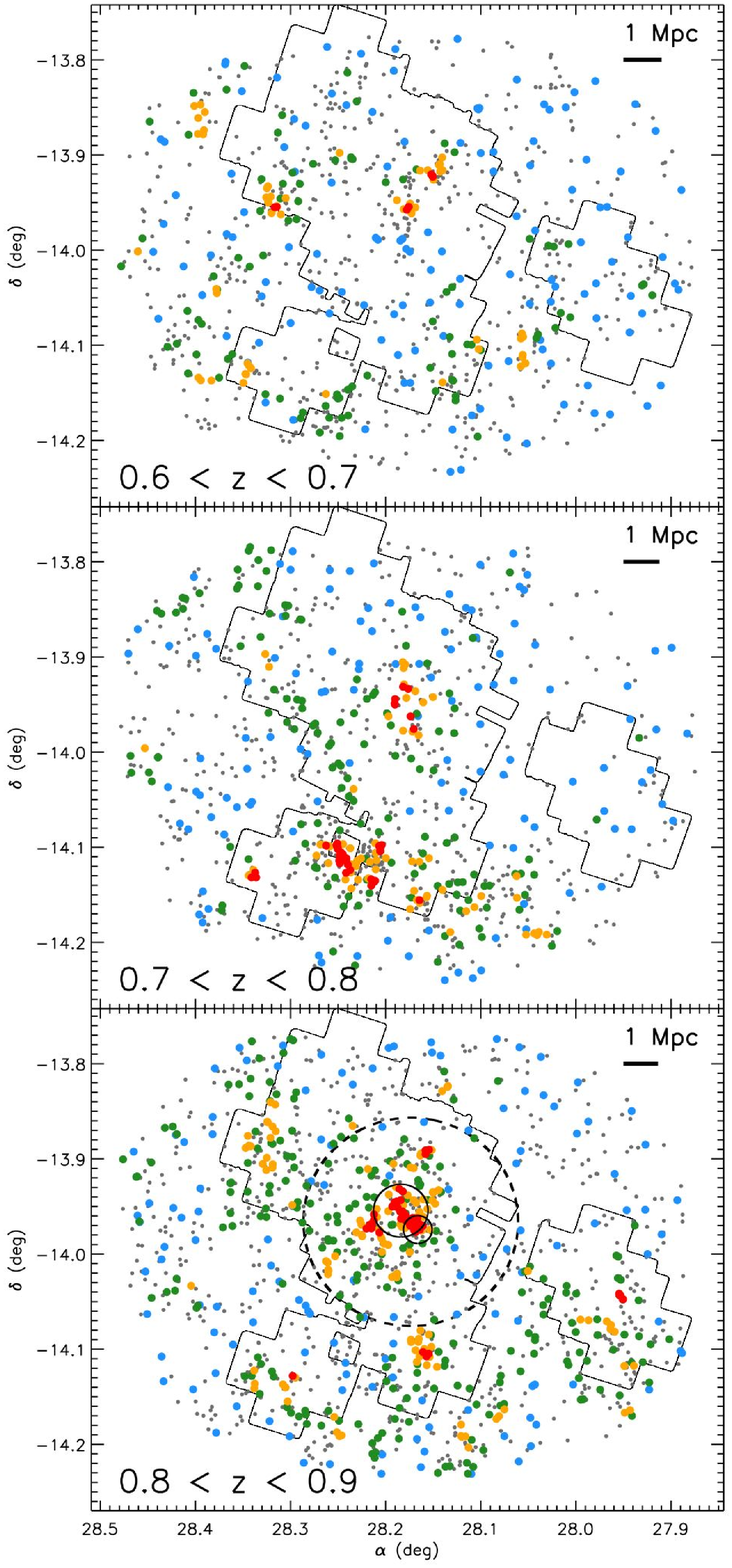}
\caption{Spatial distribution of galaxies with \zldp\ in three different redshift slices, $0.6<z<0.7$ (top), $0.7<z<0.8$ (middle), and $0.8<z<0.9$ (bottom).  Galaxies above \mcutrangelog\ are color-coded by local density, with divisions for the four density bins at 3, 13, 50~\galmpc\ indicated by the blue, green, orange, and red points.  Galaxies below this mass limit are shown in gray.  The bar at the top right of each panel indicates a projected length of 1~Mpc.  An outline of the {\em Spitzer} MIPS 24~\micron\ imaging with exposure level above $\sim 1500$~s is shown in black.  The cluster \rxj\ is at $z \sim 0.83$ and is indicated by the two solid black circles that represent the virial radii of its two cores.  The dashed circle represents a projected clustercentric radius of 3~Mpc.} \label{densitymap}
\end{figure}

We compute the projected local galaxy density using a similar but slightly updated procedure as carried out in \citet{patel2009b}.  For a given galaxy at redshift $z_{0}$, we determine the distance to the 5th nearest neighbor, $d_5$, where only those galaxies above the mass-limit of \mcutrangelog\ and with $|z_{\rm neigh} - z_{0}|/(1+z_{0}) < 2\%$ are included in the neighbor list.  Note that this velocity window is twice as large as the typical uncertainty in $(1+z_{\rm LDP})$.  In determining $d_5$, distances to all galaxies in the neighbor list are sorted from lowest to highest and the corresponding weights (computed from Figure~\ref{fig_compmap} in the Appendix) of the neighbors from the completeness map are summed.  We interpolate to find the distance corresponding to a sum of five in the weights (i.e., $d_5$).  In this way, our densities reflect values for a survey with 100\% spectroscopic completeness.  The distance, $d_5$, is used to define the circular areal element for computing the local density.  Thus, the local density is $\Sigma = 5/(\pi d_5^2)$.  The analysis in this paper is limited to the central regions of our spectroscopic coverage (i.e., where there is $K_s$ or MIPS imaging), therefore minimizing edge effects.  In addition, the list of neighbors for objects near the edge of the redshift window containing our sample includes objects that lie slightly beyond the redshift boundaries.  The median density of our full mass-limited sample at \zwindowwide\ with $K_s$-band imaging is $\sim 6.5$~\galmpc.  Ignoring galaxies in the redshift interval containing the cluster, \zwindow, the median density is $\sim 4.8$~\galmpc.

Figure~\ref{densitymap} shows galaxies above the mass limit in three redshift slices and color coded by local density.  The cluster \rxj\ is at a redshift of $z \sim 0.83$ and is easily seen in the center of the bottom panel.  Note how this redshift slice, $0.8<z<0.9$, contains a significant amount of structure compared to the two lower redshift slices.  Another prominent overdensity is the large structure at $z \sim 0.75$.  We note that \citet{tanaka2006} identify it as a cluster but measure a velocity dispersion for this structure of only $210 \pm 98$~\kmps\ from a sparse sample of objects.  The nature of this overdensity is therefore unclear.

\subsection{Star Formation Rate Indicators} \label{sfr_sec_sfr}

We use three different measures of the SFR in this paper: (1) rest-frame 12-15~\micron\ luminosities from {\em Spitzer} MIPS 24~\micron\ imaging, (2) the SFR of the best-fit model from the SED fitting, and (3) \oii\ line luminosities.  Each of these SFR indicators has associated with it various systematics \citep[see e.g.,][]{kennicutt1998}.  It is therefore helpful to use multiple indicators in drawing conclusions about the SF activity of galaxies in different environments.

\subsubsection{Spitzer MIPS 24~\micron}

The mid-IR emission traced by MIPS in our redshift interval of study (rest-frame $12-15$~\micron\ for our sample) represents a combination of thermal emission from dust grains that reprocess UV light as well as emission lines from polycyclic aromatic hydrocarbons \citep[PAHs, see review by][]{puget1989}.  This mid-IR luminosity correlates with the total infrared luminosity, $L_{\rm IR}$ \citep[$8-1000$~\micron,][]{chary2001}, which serves a tracer for the total amount of re-emitted UV-visible light.  Typical conversions between $L_{\rm IR}$ and SFR are calibrated with starbursts that have ages of $<10^8$~yr \citep{kennicutt1998}.  However, the SF timescales probed by $L_{\rm IR}$ can vary depending on the heating source (e.g., young vs. old stars) and optical depth of the dust.  Some works find that MIPS derived SFRs may represent SF over long timescales \citep[$\sim 1-2$~Gyr, see, e.g.,][]{salim2009,kelson2010} given the non-negligible contribution of dust heating associated with intermediate age stellar populations and TP-AGB stars.  This has the effect of overestimating the amount of SF at the epoch of observation given that studies of the SF history of the universe \citep{lilly1996,madau1996c} indicate that SFRs were higher at earlier epochs.  Meanwhile, recent Herschel observations indicate that the observed MIPS 24~\micron\ flux by itself underestimates the actual $L_{\rm IR}$ at these redshifts when compared to $L_{\rm IR}$ derived from the PACS 100~\micron\ and 160~\micron\ bands \citep{rodighiero2010b}.  The lack of consensus in the literature with regards to SFR measurements derived solely from MIPS 24~\micron\ data is clear.  For this reason, we focus on the {\em relative} changes in the MIPS derived SFRs across different environments.

We follow the method of \citet{wuyts2008} and others \citep[e.g.,][]{damen2009} in deriving a total infrared luminosity, $L_{\rm IR}$, from the MIPS data.  This technique uses the IR templates of \citet{dale2002} to convert the observed 24~\micron\ flux into $L_{\rm IR}$.  For a given redshift, $L_{\rm IR}$ is computed for IR templates spanning a range of heating levels of the interstellar environment.  The mean of all $\log(L_{\rm IR})$ is used to represent the final $L_{\rm IR}$.  We note that the derivation of $L_{\rm IR}$ in this work differs from that in \citet{patel2009b}.  To convert $L_{\rm IR}$ into a SFR we use Equation (4) from \citet{kennicutt1998}, but adjusted for a Chabrier IMF by multiplying by a factor of $0.56$.  While AGN also emit at these mid-IR wavelengths, they do not contribute significantly to the overall mid-IR luminosity density at these redshifts \citep{bell2005}.  Their contribution to the mid-IR flux for individual galaxies is also not very high \citep{salim2009}.

In this work, \nmipsks\ galaxies at \zwindowwide\ with $K_s$ imaging and stellar masses \mcutrange\ have MIPS coverage.  The depth of the MIPS imaging precludes a galaxy-by-galaxy analysis as only $\sim 28\%$ of galaxies above the mass limit are detected above $75$~\uJy.  In addition, this detection rate varies for galaxies in different stellar mass and local density ranges.  As a consequence, it is not possible to characterize the distribution of individual MIPS fluxes for a given sample with a mean or median, as many galaxies are undetected.  In order to compute the median 24~\micron\ flux for a given sample, we therefore stacked MIPS imaging for all galaxies in the sample, combining both detected and undetected sources.  This approach allowed us to carry out a uniform analysis across various sub-samples (e.g., mass and density) and enabled us to reach flux levels below the MIPS detection limit.  To compute the median MIPS flux and its uncertainty for a given stack, we carry out a bootstrap analysis as follows.  For a given sub-sample of $N$ galaxies, we extract postage stamps from the background subtracted MIPS imaging, centered at the RA and Dec of each galaxy.  We draw $N$ random postage stamps, with replacement, taking into account the weights from the completeness map, and compute the median of the stack at each pixel position, thus creating a median image from the $N$ postage stamps.  Aperture photometry is performed on the median image using an aperture of diameter $D=6\arcsec$.  The resulting flux is aperture corrected and the value stored.  This process is repeated 1000 times.  The mean and standard deviation of the 1000 measurements for the median are taken to represent the median flux and its uncertainty.  These flux values are then converted into a SFR using the median redshift of the sample to determine $L_{\rm IR}$.  This stacking technique was also employed in \citet{patel2009b}.  In this work, we build on the analysis from \citet{patel2009b} by expanding our sample redshift range from \zwindow\ to \zwindowwide.

We compute the median of the stack rather than the mean in order to minimize the contribution from neighboring bright sources. We note that using the mean would have resulted in fluxes that were roughly $0.2$~dex larger than what we found for the median.  However, the final results would have been qualitatively the same.

In order to test the robustness of the median flux estimates from the stacking analysis, we stacked MIPS imaging of detected objects and found the median stacked flux to agree with the median of the individual flux measurements to within $<10\%$.  This was also true for detected MIPS objects in the highest density regions of our sample.

\subsubsection{Star Formation Rates from SED Fit}

\begin{figure*}
\epsscale{1.2}
\plotone{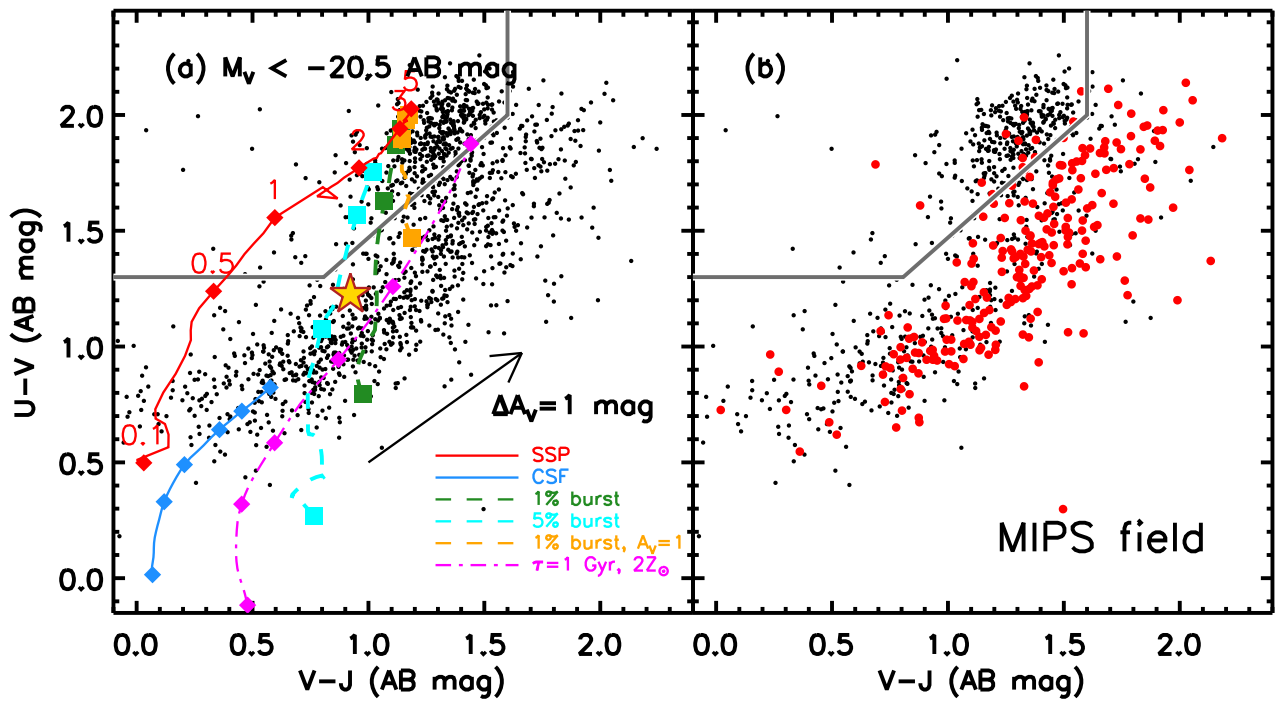}
\caption{Rest-frame $U-V$ versus $V-J$ for galaxies at \zwindowwide.  (a) $UVJ$ diagram for a luminosity limited sample, $M_V < -20.5$~AB mag ($L_V \sim 1.35\times10^{10}$~\lsun).  The \citet{williams2009} boundary separating quiescent (upper left) and SFGs (lower right) is marked by the gray wedge.  The arrow indicates the reddening vector for $\Delta A_V =1$~mag.  Evolutionary tracks are shown for various BC03 solar metallicity SFHs.  SSP (red) and CSF (blue) models are shown with time steps indicated by the diamonds at $t=~$0.1, 0.5, 1, 2, 3, 5 Gyr.  After $\sim 3-5$~Gyr, the SSP falls near the quiescent clump, while the CSF lies near very blue SFGs.  The colors of SFGs are extended to redder colors with the addition of dust.  Green and cyan dashed lines represent the evolution of $1\%$ and $5\%$ bursts added to a $t=4$~Gyr SSP and followed at time steps after the burst, $t_{\rm AB}$, of $t_{\rm AB}=~$0.01, 0.1, 0.3, 0.5~Gyr (marked by squares; $t_{\rm AB}$ increases from bottom to top).  The orange dashed line represents the $1\%$ burst scenario but with the burst component extincted by $A_V=1$~mag.  The yellow star represents low levels of continuous SF (e.g., analogous to the Milky Way) with a CSF that has been ongoing for the last $1$~Gyr combined with an SSP of $t=4$~Gyr.  The CSF component contributes $5\%$ of the total mass.  The magenta dash-dot line represents a twice solar metallicity $\tau$-model with $\tau=1$~Gyr and the same time steps as the SSP and CSF marked with diamonds.  In general, galaxies with any recent SF activity will be classified as SFGs, while those without will be classified as quiescent. (b) The subset of $M_V < -20.5$~AB mag galaxies with MIPS coverage.  Those with MIPS detections ($>75$~\uJy) are indicated in red and fall primarily in the region occupied by SFGs. \label{bicolor}}
\end{figure*}

Each BC03 $\tau$-model that is fit to an SED has associated with it an instantaneous SFR.  The SFRs derived from the SED fitting primarily leverage the rest-frame UV continuum, sampled by the observed $BV$ filters.  Stars above $\sim 5$~\msun\ dominate the light output at these wavelengths and the SFR measurements are valid for timescales of $10^8$~yr \citep{kennicutt1998}.  Because dust extinction is included in the $\tau$-models, the SED SFRs represent an extinction corrected SFR.  We caution that degeneracies in the parameters of the SED fitting often lead to a range of SFR values for a single galaxy, often dependent on the value of \av.  The typical uncertainty for \av\ is $0.3-0.4$~mag and was computed by refitting SEDs for a subset of our mass-limited sample after injecting Gaussian noise into the observed flux measurements and comparing the resulting values for \av\ with the original ones.  We use the SFR from the best fitting $\tau$-model to represent the SED SFR.

\subsubsection{\oii\ Emission Line Luminosities}

Nebular emission lines serve as a tracer for the SFR as they re-emit UV light blueward of the Lyman limit produced by stars with masses of $>10$~\msun\ and lifetimes of $\sim 20$~Myr \citep{kennicutt1998}.  As part of the SED fitting procedure, we include a Gaussian component for \oii\ (see Figure~\ref{sfr_fig_sedfits}), a prominent SFR tracer that lies within the LDP wavelength range for our redshift interval of study.  The Gaussian \oii\ component is normalized such that the coefficient in the NNLS fit to [OII] represents the observed line flux, $F_{\rm [OII]}$ (units: erg~s$^{-1}$~cm$^{-2}$).  This line flux is converted into a luminosity and then a SFR using Equation~3 from \citet{kennicutt1998}, and scaled by a factor of $0.56$ for a Chabrier IMF.

We estimate extinction corrected [OII] luminosities by utilizing the best fitting \av\ from the SED fit and a \citet{calzetti2000} extinction law.  We assume $E(B-V)_{gas} = E(B-V)_{stars}/0.44$, as indicated in \citet{calzetti2000}.  We caution that these extinction corrections have large uncertainties given the uncertainty in \av\ (see previous section).

In the sections to follow, we discuss SED and [OII] derived SFRs for the \nmcut\ galaxies at \zwindowwide\ with $K_s$ imaging and stellar masses above \mcutrangelog.  Approximately $\sim 82\%$ of galaxies in this sample have a non-zero [OII] flux from the NNLS fit.  Note that this sample is larger than the MIPS sample.

\section{Galaxy Classification with a Color-Color Diagram: Distinguishing Quiescent and Star Forming Galaxies} \label{sfr_sec_bicolor}

One of the main science goals of this paper is to determine how the SFRs of galaxies vary across a range of environments.  In particular, we would like to know how SFGs contribute to the SSFR-density relation.  In \citet{patel2009b} we hypothesized that the change in the median SFR with density was driven simply by a changing proportion of quiescent and SFGs, as determined from the fraction of red galaxies, leaving a negligible contribution from any change in the SFRs of SFGs with density.  Many other works also routinely use the red-sequence, as defined with a rest-frame color and magnitude (or stellar mass), to represent quiescent galaxies \citep[e.g.,][]{bell2004b,faber2007,brown2007}.  However, dust can lead to a significant fraction of SFGs occupying the red-sequence \citep{vanderwel2007b,maller2009,whitaker2010}.  Below, we utilize a color-color selection that allows us to identify these reddened SFGs, in addition to blue SFGs, and distinguish both types from quiescent galaxies.

\subsection{The $UVJ$ Diagram}

Figure~\ref{bicolor} shows rest-frame $U-V$ versus $V-J$ (hereafter referred to as the $UVJ$ diagram).  Such color-color diagrams have been utilized by several other works to distinguish SFGs from quiescent galaxies \citep{wuyts2007,williams2009,williams2010,wolf2009,balogh2009,whitaker2010}.  While $U-V$ primarily identifies age, $V-J$ is sensitive to age {\em and} reddening owing to the long baseline in wavelength.  One key feature of the $UVJ$ diagram is the red ``quiescent clump'' ($U-V \sim 2$~AB mag and $V-J \sim 1.3$~AB mag), the sub-sample of red-sequence galaxies that lack SF, as shown later.  Below the quiescent clump lies a sequence of SFGs.  We use the color boundaries derived by \citet{williams2009} to separate quiescent galaxies from SFGs.  Model BC03 evolutionary tracks are shown for an SSP (red line) and ``constant'' SFR (CSF, blue line) model ($\tau$-model with $\tau=20$~Gyr) for solar metallicity.  The SSP model track passes near the quiescent clump after $t=3$~Gyr, while the CSF model track terminates at the blue end of where SFGs lie.  The reddening vector in Figure~\ref{bicolor}$a$ shows that SFGs are extended to redder colors with the addition of dust extinction.  Note that some of the reddest SFGs ($V-J \gtrsim 1.6$~AB mag) would have been classified as being ``red-and-dead'' based on $U-V$ colors alone.  Instead, these galaxies are distinct from the quiescent clump and are predicted to be SFGs with high values of \av.  Bicolor diagrams, such as the $UVJ$ ones shown here, can therefore be used to identify SFGs, including those that are reddened.

\subsection{A Diversity of Star Formation Histories for $UVJ$-selected SFGs}

While the SSP and CSF model tracks in Figure~\ref{bicolor}$a$ provide a glimpse of how simple galaxy SFHs appear in the $UVJ$ diagram, the actual SFHs of galaxies are likely to be more complex.  In Figure~\ref{bicolor}$a$ we also show how a variety of SFHs would appear in the $UVJ$ diagram and emphasize the effectiveness of the \citet{williams2009} boundary in distinguishing quiescent and SFGs.  These examples also demonstrate that SFGs need not have formed the bulk of their mass recently in order to be classified as an SFG.  Even the slightest amount of recent SF can move formerly quiescent galaxies into the SFG region of the $UVJ$ diagram.

SFHs that proceed with several ongoing bursts have been suggested as a mode for galaxies to form a large fraction of their stellar mass, especially at these higher redshifts \citep{bell2005,dressler2009b}.  The trajectories of various bursting SFHs are shown in Figure~\ref{bicolor}$a$ as dashed lines.  The green and cyan colored tracks show the evolution of a $1\%$ and $5\%$ burst (by mass) added to a $4$~Gyr solar metallicity SSP at time steps after the burst, $t_{\rm AB}$, of 0.01, 0.1, 0.3, and 0.5~Gyr (time steps shown as squares).  Immediately after the burst (i.e., $t_{\rm AB} = 0.01$~Gyr), the $U-V$ colors become bluer then most SFGs in our sample.  After $t_{\rm AB} \approx 0.1-0.3$~Gyr, the colors once again return to resemble a galaxy residing in the quiescent clump.  This transition proceeds more quickly for the smaller $1\%$ burst.  These models also demonstrate that {\em unreddened} bursts on top of old stellar populations are likely not capable of producing SFGs with the reddest colors in the $UVJ$ diagram.  Those red SFGs are most likely to be obscured at some level.

The orange track in Figure~\ref{bicolor} represents a $1\%$ burst added to an SSP but with the burst component extincted by \av$~=1$~mag.  Note the displacement from the unobscured $1\%$ burst (green track). These dusty bursts spend even less time ($<0.1$~Gyr) classified as SFGs.  Depending on the level of extinction and the distribution of the dust, reddened bursts can populate the reddest colors for SFGs.

A galaxy with low levels of continuous SF, qualitatively resembling the state of the Milky Way, is represented by the single yellow star at $U-V \sim 1.2$~AB mag and $V-J \sim 0.9$~AB mag.  This SFH is a combination of a CSF that has been ongoing for the last 1~Gyr and an SSP of age 4~Gyr.  The CSF component has produced $5\%$ of the total mass.  Such low levels of continuous SF would place a galaxy in the region occupied by SFGs.

SFHs with SFRs that slowly decline with time can lead to galaxies being classified as SFGs despite very low SSFRs, especially for super-solar metallicity stellar populations.  The magenta dash-dot track represents a twice solar metallicity $\tau$-model with $\tau=1$~Gyr.  Time steps similar to that of the SSP and CSF are shown for this model.  The use of a super solar metallicity model generally results in redder $V-J$ colors at fixed $U-V$.  After $3-5$ $e$-folding times in the SFR (i.e., between the last two time steps), this $\tau$-model would remain classified as a SFG with a SSFR between $\sim 0.1$ and $0.02$~Gyr$^{-1}$.  We note that for the corresponding solar metallicity SFH (not shown), the model track would cross the boundary between SFG and quiescent after $\sim 3$~Gyr.  Thus, some of the galaxies in the ``green valley'' ($1.2 \gtrsim U-V \gtrsim 1.6$~AB mag) are likely to be comprised of galaxies with somewhat older stellar populations with some ongoing SF, rather than SFGs with more vigorous SF and modest amounts of dust (e.g., $A_V \sim 1$~mag).

\begin{figure}
\epsscale{1.2}
\plotone{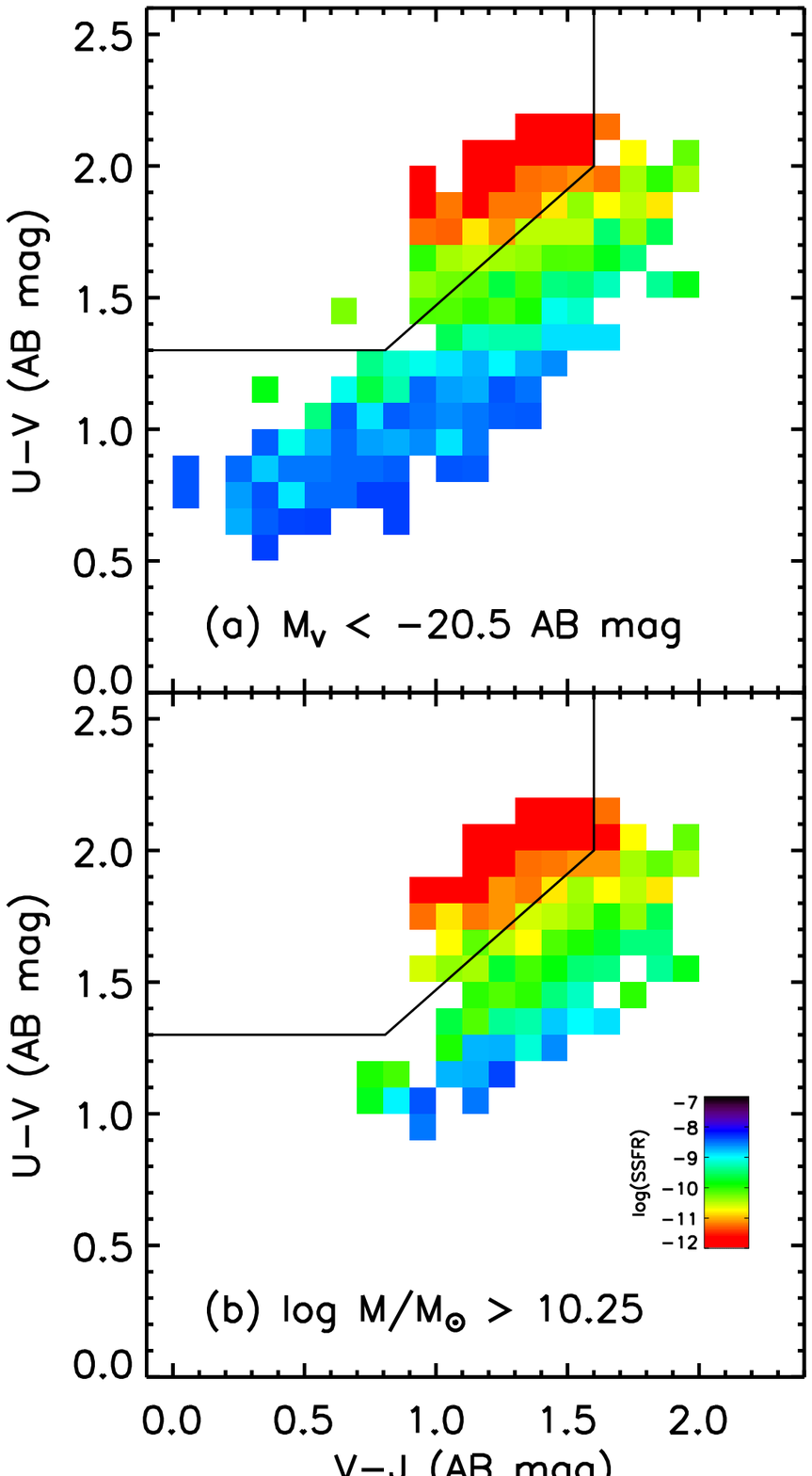}
\caption{$UVJ$ diagram color-coded by the median SED SSFR in color-color bins of size $0.1$~mag. (a) Galaxies at \zwindowwide\ with rest-frame $M_V < -20.5$~AB mag. (b) Galaxies above a stellar mass of \mcutrangelog.  The colors that correspond to different $\log$(SSFR) are shown in the bottom right.  Bins with fewer than three objects are not included in the analysis.  The \citet{williams2009} boundary separating quiescent galaxies from SFGs is shown in black.  Note how selecting by mass minimizes the bias against faint red galaxies and therefore results in fewer blue SFGs compared to the luminosity selection in (a). \label{sfr_fig_UVJ_2d}}
\end{figure}

\begin{figure*}
\epsscale{1.1}
\plotone{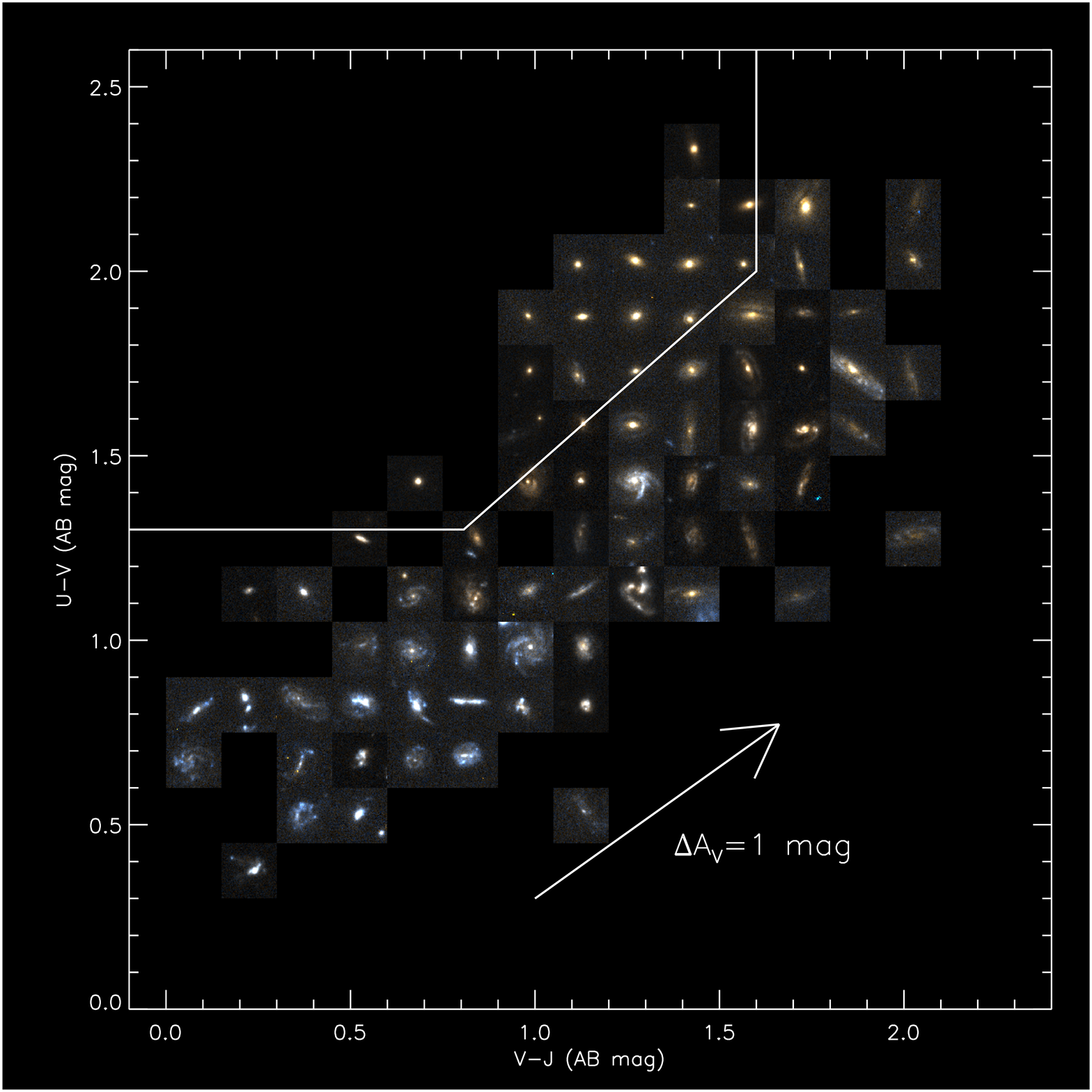}
\caption{$UVJ$ diagram represented by HST ACS postage stamps ($3\arcsec$ on a side) of galaxies at \zwindowwide\ with $M_V < -20.5$~AB mag.  The \citet{williams2009} boundary separating quiescent (top left) and SFGs (bottom right) is marked in white.  Quiescent galaxies are dominated by bulges.  The bluest SFGs ($V-J \lesssim 0.8$~AB mag) are primarily disks, while the reddest SFGs ($V-J \gtrsim 1.6$~AB mag) appear to be inclined disks. \label{uvj_hst}}
\end{figure*}

To summarize, based on BC03 stellar population synthesis models, galaxies classified as star-forming in the $UVJ$ diagram can be comprised of (1) galaxies with relatively constant SFRs since the onset of SF, (2) formerly quiescent galaxies that have undergone a very recent burst, (3) galaxies with some residual, low level, long-term SF (e.g., $5\%$ of mass added over the last $1$~Gyr with roughly constant SFR), and (4) super-solar metallicity galaxies with low levels of SF.  The addition of dust to any of these scenarios moves SFGs towards redder $U-V$ and $V-J$ colors, and generally keeps them in the region occupied by SFGs.  Thus, galaxies with virtually any level of recent SF activity would be classified as SFGs, while those without would be classified as quiescent based on the \citet{williams2009} boundary.

\subsection{MIPS Detections Trace $UVJ$-Selected SFGs}

\begin{figure*}
\epsscale{1.2}
\plotone{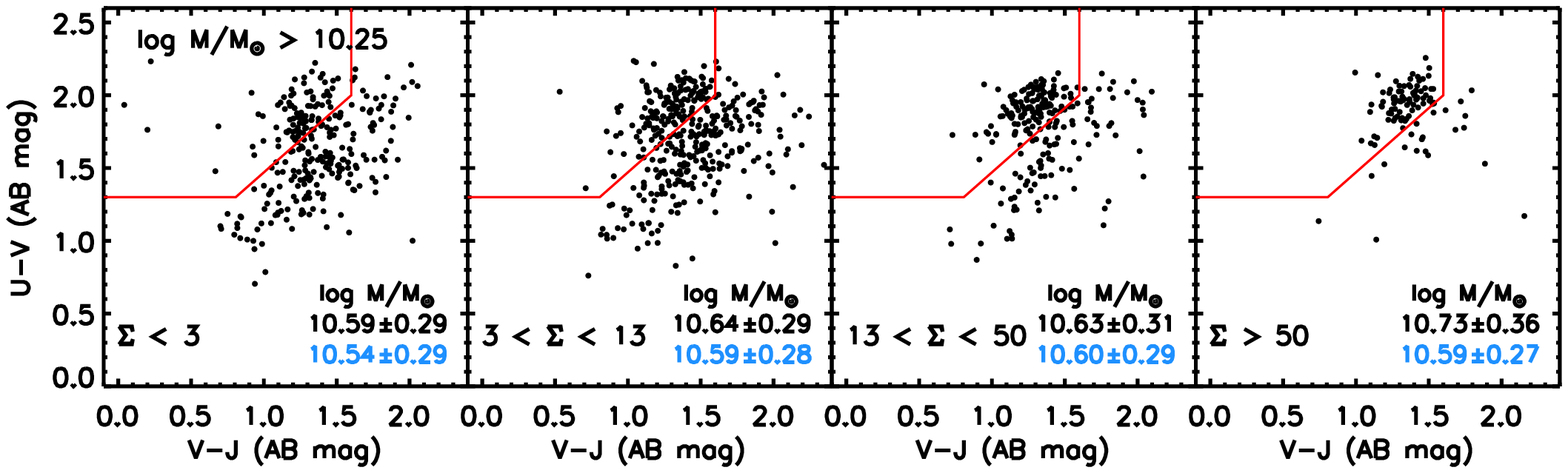}
\caption{UVJ diagram for galaxies at \zwindowwide\ with mass \mcutrangelog\ in four density bins.  The red line denotes the \citet{williams2009} boundary separating quiescent and SFGs.  Note that selecting galaxies above a mass limit of \mcutrangelog\ results in fewer blue SFGs compared to the luminosity-limited selection in Figure~\ref{bicolor}.  The median and standard deviation of $\log M/M_{\odot}$ for the full mass-limited sample (black) and of the SFG sample (blue) are labeled in the bottom right for each density bin.  The mass function is fairly uniform in different environments for both samples, with slightly more massive galaxies at the highest densities.  Mass dependent correlations for galaxy properties in different density bins are therefore minimized by using a stellar mass limited sample.} \label{sfr_fig_UVJ_density}
\end{figure*}

The models discussed above provide a theoretically motivated backing for the \citet{williams2009} boundary separating quiescent and SFGs.  Here we perform an empirical check on this boundary.  Figure~\ref{bicolor}$b$ shows galaxies with MIPS $24$~\micron\ coverage and indicates the subset with $24$~\micron\ detections ($>75$~\uJy).  These detections primarily trace the region occupied by SFGs.  Note the detections at red colors ($U-V>1.6$~AB mag) where galaxies would have been considered members of the red-sequence based solely on their $U-V$ color.  Most of these MIPS detections are in the region of the $UVJ$ diagram consistent with SFGs with extinction.  A stacking analysis of the SFGs in this region ($U-V>1.6$~AB mag) {\em without} MIPS detections suggests that they are slightly below the detection limit with a median 24~\micron\ flux of $\sim 41$~\uJy, and therefore have modest obscured SF activity as well ($\sim 3$~\sfr\ at $z=0.75$).  In contrast, the median stacked MIPS 24~\micron\ flux for quiescent galaxies at $U-V>1.6$~AB mag is a factor of $\sim 3$ lower.  The MIPS detections therefore lend further support to our ability to isolate SFGs in the $UVJ$ diagram, including both blue (i.e., unattenuated) and red (i.e., dust reddened) SFGs.

\subsection{The Distribution of Star Formation in the $UVJ$ Diagram}

The models tracks shown in Figure~\ref{bicolor} provide clues on the recent SFHs of galaxies that occupy different regions of the $UVJ$ diagram.  However, it is helpful to know how the SFRs or SSFRs of galaxies in our sample are distributed in the diagram, thus providing a check on our division between quiescent and SFGs.  Figure~\ref{sfr_fig_UVJ_2d} shows the $UVJ$ diagram divided into color-color bins of size $0.1$~mag, with each bin color-coded by the median SSFR determined from the SED fit.  The top panel shows a luminosity-selected sample of galaxies at \zwindowwide\ with $M_V<-20.5$~AB mag, while the bottom panel shows a mass-selected sample with \mcutrangelog.  Note that selecting galaxies above a stellar mass limit of \mcutrangelog\ minimizes the bias against faint red galaxies and therefore results in fewer blue SFGs compared to a luminosity selection.  Lines of constant SSFR span from blue $U-V$ and $V-J$ colors to redder ones for SFGs.  Thus, SFGs with red $U-V$ and $V-J$ colors indicative of high levels of reddening, perhaps owing to their inclination, can have similar SED based SSFRs to those with bluer colors.  At a fixed $V-J$ color, SFGs with lower SED based SSFRs have redder $U-V$ colors.  The recent SFH of a galaxy therefore also determines where SFGs lie in the $UVJ$ diagram, in addition to the level of extinction.  Figure~\ref{sfr_fig_UVJ_2d} shows that the \citet{williams2009} boundary between quiescent and SFGs is well justified.  Note too that although quiescent galaxies with bluer $U-V$ colors ($U-V<1.7$~AB mag) have somewhat higher SSFRs than redder ones, there are very few galaxies in that region of $UVJ$ color-space (see Figure~\ref{bicolor}$a$).

Finally, it is worth noting that while there appears to be a strong gradient at a fixed $V-J$ color in SED based SSFRs for SFGs, \citet{williams2009} find a much smaller range of SSFRs for MIPS based SFRs.  \citet{williams2010} however also find a stronger gradient for SED based SSFRs.  These differences between SFR indicators are important to consider when interpreting the results below, especially for the SFRs of SFGs (see \S~\ref{sec_ssfr_sfg}).

\subsection{A Morphological Overview of the $UVJ$ Diagram}

Figure~\ref{uvj_hst} populates the $UVJ$ diagram with HST ACS color postage stamps for a luminosity-limited sample.  These high resolution images provide a sense for how disk or bulge dominated systems, and their orientations, help to explain what is seen in different regions of the $UVJ$ diagram.  The quiescent region is dominated by galaxies with strong bulge components while the star-forming region is comprised of disks.  Blue SFGs appear to be face-on disks and the bluest of them ($U-V \sim 0.7$~AB mag) look like bursting systems -- consistent with the location in $UVJ$ color space predicted by the bursting tracks discussed in Figure~\ref{bicolor}$a$.  Meanwhile, the reddest SFGs appear to be edge-on disks, consistent with the red colors arising from extinction.  Interestingly, several SFGs with intermediate colors ($U-V \sim 1.6$~AB mag and $V-J \sim 1.4$~AB mag) appear to have clear disks but also host a strong bulge component.  The colors of these systems likely arise from older stellar populations, perhaps in a transition phase, rather than dust.  A more detailed discussion of the morphological structure of galaxies in various environments and in the $UVJ$ diagram will be presented in a future paper (Patel et~al. 2011, in preparation).

\subsection{The $UVJ$ Diagram in Different Environments}

Our analysis in \S~\ref{sfr_sec_sfrdensity} utilizes the $UVJ$ diagram in different density bins in order to examine relative proportions of quiescent and SFGs as well as the SFRs of SFGs.  Figure~\ref{sfr_fig_UVJ_density} shows the $UVJ$ diagram in four density bins for galaxies at \zwindowwide\ above the mass limit of \mcutrangelog.  Note the presence of many galaxies with red $U-V$ colors ($U-V \gtrsim 1.7$~AB mag) that are classified as SFGs.  For galaxies above the mass limit that would be classified as red-sequence members based solely on their $U-V$ colors \citep[based on a red-blue separation method similar to that applied in][]{patel2009}, roughly $\sim 32\%$ are classified here as $UVJ$-selected SFGs.  \citet{balogh2009} find a similar value at somewhat lower redshifts and for a slightly lower mass limit.  Thus, a considerable portion of optically red galaxies in our redshift interval are dusty and star-forming.

Figure~\ref{sfr_fig_UVJ_density} also indicates the median and standard deviation of the logarithm of the stellar mass in different density bins for our galaxy sample above the mass limit of \mcutrangelog\ (black), as well as for the sample of SFGs (blue).  The median stellar mass varies by less than $\lesssim 0.15$~dex between different density bins for both samples.  The scatter in stellar mass between density bins is also comparable.  Given that the distribution of stellar masses in different density bins are roughly similar, mass-dependent correlations with various galaxy properties cannot be solely responsible for any density-dependent correlations \citep[see also,][]{cooper2010b}.

\section{A Declining SSFR-Density Relation at \zwindowwide} \label{sfr_sec_sfrdensity}

\begin{figure}
\epsscale{1.2}
\plotone{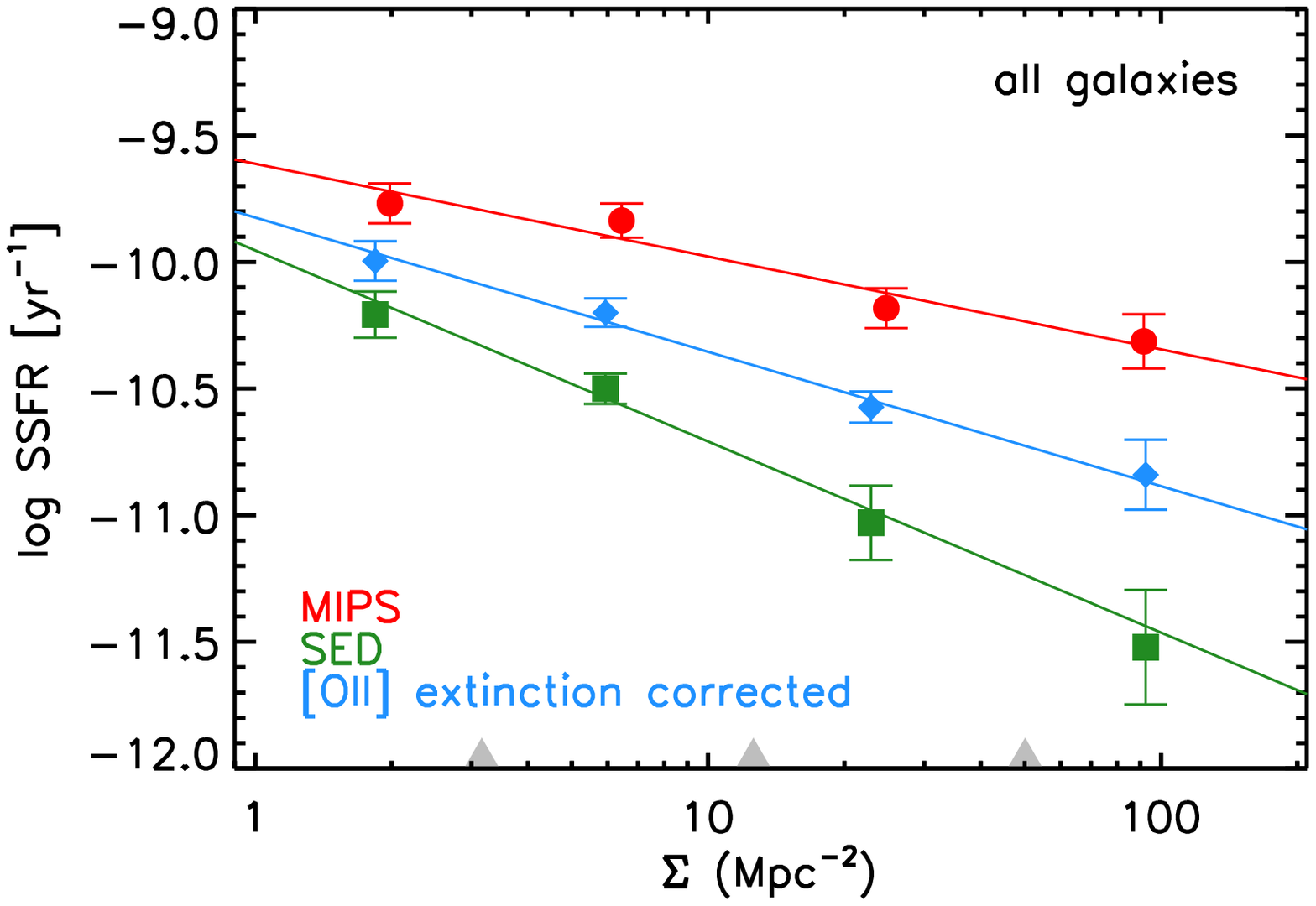}
\caption{Median SSFR versus local density for galaxies with mass \mcutrangelog\ at \zwindowwide.  Gray triangles on the $x$-axis denote boundaries of density bins.  SFRs are derived from three different indicators: (1) rest-frame 12-15~\micron\ luminosities from {\em Spitzer} MIPS 24~\micron\ imaging (red circles), (2) SED fitting (green squares), and (3) extinction corrected \oii\ emission line luminosities (blue diamonds).  The colored lines indicate the best fitting log(SSFR)-log(density) relations with corresponding parameters given in Table~\ref{table_ssfr_density}.  For all three SFR indicators the SSFR-density relation decreases, from the low-density field to the cores of groups and a cluster. \label{ssfr_density_all}}
\end{figure}

In this section, we study the SSFR-density relation for galaxies at \zwindowwide\ with stellar masses above \mcutrangelog.  We utilize the three different SFR indicators discussed in \S~\ref{sfr_sec_sfr}: (1) rest-frame 12-15~\micron\ luminosities from {\em Spitzer} MIPS 24~\micron\ imaging, (2) SED fitting, and (3) extinction-corrected \oii\ emission line luminosities.  Each of these SFR indicators has associated with them various systematic uncertainties.  For the purposes of this paper, we are interested in the {\em relative} change in the SFR within a single indicator, across different density bins.  We note that the MIPS derived SFRs are limited to the sample with MIPS coverage (see Figure~\ref{densitymap}), while the SED and [OII] SFRs are reported for a larger sample.

Figure~\ref{ssfr_density_all} shows the median SSFR-density relation for each of the three SFR indicators.  Uncertainties were computed for the SED and [OII] SSFRs by bootstrapping the sample within each density bin.  A bootstrap technique was also used to compute the uncertainties for the MIPS SSFRs (see \S~\ref{sfr_sec_sfr}).  The position of the data points along the $x$-axis reflects the median value for the local density within the given sub-sample.  At \zwindowwide, the SSFR from all three SFR indicators declines across the entire range of densities, from the low-density field to the cores of groups and a cluster.  Thus, the overall SF activity in the cores of groups and a cluster at this epoch is lower than in the field.  Straight line fits to the MIPS, SED, and [OII] log(SSFR)-log(density) relations indicate non-zero, negative slopes at significances of $\sim 5$, $6$, and $7\sigma$.  The best fit parameters are given in Table~\ref{table_ssfr_density}.  The SED and [OII] based SSFRs decline by roughly an order of magnitude across the full density range, while the MIPS based SSFRs decline by a factor of $\sim 4$.  We note that the raw [OII] SSFRs that are uncorrected for extinction (not shown), which depend on the SED fits only in establishing a continuum level above which the line flux is integrated, also show a declining SSFR with increasing density, although with a shallower slope ($\Delta(\log$~SSFR)/$\Delta(\log \Sigma) \approx -0.35 \pm 0.05$).  This suggests relatively higher levels of extinction at lower densities (see \S~\ref{sec_av_density}).

A large proportion of our galaxy sample in the highest density environments reside in the vicinity of the cluster.  Some of these galaxies have likely passed through the cluster core at some point in time, resulting in their lower SFRs \citep[see e.g.,][]{balogh2000}.  It is therefore reasonable to ask if the cluster is solely responsible for driving the declining SSFR-density relation that we found in Figure~\ref{ssfr_density_all}.  Figure~\ref{ssfr_density_out_all} shows the mass-limited SSFR-density relation for all three SFR indicators after removing galaxies near the cluster with redshifts in the range \zwindow.  Removing galaxies in this redshift range primarily impacts the two highest density bins (note the larger error bars).  The SSFR-density relation continues to decline at higher densities for all three SFR indicators, although the decline is not as steep for the MIPS based SFRs.  Straight line fits to the MIPS, SED, and [OII] log(SSFR)-log(density) relations yield non-zero, negative slopes at the $\sim 2.5$, $5$, and $5\sigma$ levels respectively (see Table~\ref{table_ssfr_density} for best fit parameters).  Within the uncertainties, the SSFRs of galaxies far from the cluster in the highest density bin are essentially the same as when including the cluster.  We note that we arrive at the same conclusion when also excluding the structure at $z \sim 0.75$.  This demonstrates that our result of a declining SSFR-density relation in Figure~\ref{ssfr_density_all} is general to a range of environments and not solely driven by galaxies near a rich cluster.  Galaxies in groups at these redshifts therefore also have lower median SSFRs compared to galaxies in the field.  In the sections that follow, we study the full mass-limited sample at \zwindowwide\ but emphasize that the results are qualitatively similar when using a sample that ignores the cluster redshift slice.

\subsection{Does Environment Matter or is Stellar Mass Driving the Declining SSFR-Density Relation?}

\begin{figure}
\epsscale{1.2}
\plotone{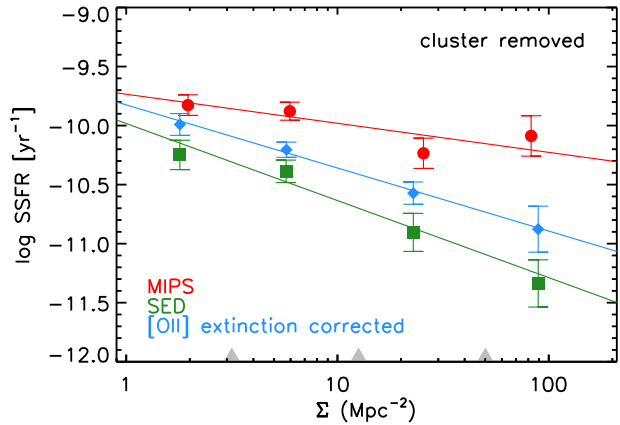}
\caption{Median SSFR versus local density for galaxies with mass \mcutrangelog\ at \zwindowwide, but only for galaxies far from the cluster \rxj\ (outside of \zwindow).  The SSFR-density relation continues to decline, indicating that galaxies near the cluster environment are not solely responsible for the trend.  Galaxies in groups at these redshifts also have lower SSFRs compared to the field. \label{ssfr_density_out_all}}
\end{figure}

Higher density environments are found to host more massive galaxies \citep[e.g.,][]{kauffmann2004,baldry2006}.  Because SFRs and SSFRs are correlated with stellar mass, it is important to account for differences in the distribution of stellar masses between environments in order to determine whether the environment plays an independent role from stellar mass in impacting galaxy SFRs.  Utilizing a simple mass cut, as shown in Figures~\ref{ssfr_density_all} and \ref{ssfr_density_out_all}, to first order does a good job of minimizing the differences in the stellar mass distributions between environments (see numbers in Figure~\ref{sfr_fig_UVJ_density}).  Here, we go a step further and study the SSFR-density relation at a fixed mass by dividing our sample into three mass bins.

\begin{figure}
\epsscale{1.1}
\plotone{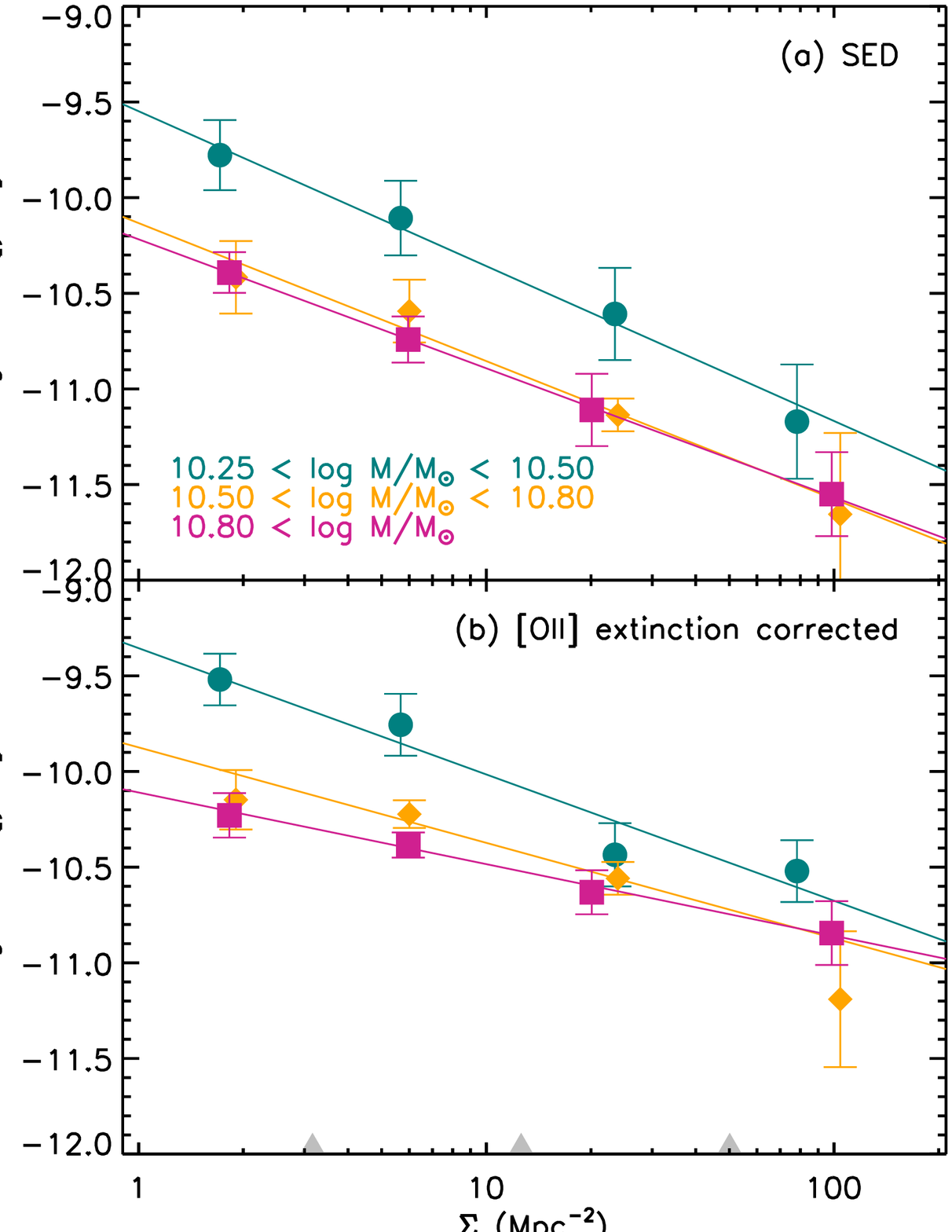}
\caption{Median SSFR versus local density for galaxies at \zwindowwide\ in three mass bins (low -- teal circles, intermediate -- orange diamonds, high -- purple squares) containing roughly equal numbers of galaxies. (a) SFRs derived from SED fitting. (b) SFRs derived from extinction corrected \oii.  The colored lines indicate the best fitting log(SSFR)-log(density) relations with corresponding parameters given in Table~\ref{table_ssfr_density_all_massbin}.  At a fixed stellar mass, the SSFR declines from low to high density for both SFR indicators by roughly an order of magnitude.  \label{ssfr_density_all_massbin}}
\end{figure}

Figure~\ref{ssfr_density_all_massbin} shows the median SSFR versus local density for all galaxies at \zwindowwide\ in three mass bins for SFRs derived from SED fitting and extinction corrected \oii.  We note that a similar figure for the MIPS based SFRs was presented in \citet{patel2009b} and arrived at qualitatively similar conclusions as presented for the other two SFR indicators below.  The mass bins in Figure~\ref{ssfr_density_all_massbin} are the same for each SFR indicator and were partitioned such that for the full sample with $K_s$ imaging, the number of galaxies in each bin was roughly the same.  The low mass bin ranges from $10.25 < \log M/M_{\odot} < 10.50$, the intermediate mass bin from $10.50 < \log M/M_{\odot} < 10.80$, and the high mass bin from $\log M/M_{\odot} > 10.80$.  The widths of the mass bins, especially the two lowest, are fairly narrow, allowing one to investigate the SSFR-density relation at a fixed mass.  For both SFR indicators and for all mass bins, the SSFR-density relation declines, from the low density field to the cores of groups and a cluster.  Table~\ref{table_ssfr_density_all_massbin} indicates the best fit parameters to the $\log$(SSFR)-$\log$(density) relations for each mass bin.  The relations for all three mass bins have negative slopes that are significant at $>3\sigma$.  Any variation in the stellar mass distribution within a mass bin across different environments would not be sufficient to explain the order of magnitude decline in SSFR found in Figure~\ref{ssfr_density_all_massbin}.  {\em This confirms that the local environment impacts the median SSFRs and SFRs of galaxies at \zwindowwide\ when controlling for stellar mass.}  

Figure~\ref{ssfr_density_all_massbin} also shows that for the two highest mass bins, the spread in SSFR at a fixed density is fairly small, implying that the SSFR-density relations are comparable.  Galaxies in these two mass bins lie above $\log M/M_{\odot} > 10.5$, which is similar to the characteristic mass found by \citet{kauffmann2003} at $z \sim 0$ above which galaxy properties appear similar.  Interestingly, the SSFRs for the highest mass galaxies ($\log M/M_{\odot} \gtrsim 10.8$) at \zwindowwide\ are also sensitive to their local environment, indicating that stellar mass is not solely responsible for driving their SF properties.

Finally, we note that because the SSFR-density relations are shown for narrow mass bins, the {\em SFR-density} relations for a given mass bin have slopes that are similar (in log-log space) to the corresponding SSFR-density relations.

\subsection{Quiescent Fraction at \zwindowwide\ as a Function of Density}

\begin{deluxetable}{lcc}
\tablewidth{0pc}
\tablecolumns{3}
\tablecaption{Parameters for best-fit $\log-\log$ lines for SSFR-density relation \label{table_ssfr_density}}
\tablehead{
\colhead{SFR Indicator} & \colhead{Intercept} & \colhead{Slope} \\
\colhead{} & \colhead{($\log$~SSFR)} & \colhead{($\Delta(\log$~SSFR)/$\Delta(\log \Sigma$))}
}
\startdata
\cutinhead{All galaxies (Figure~\ref{ssfr_density_all})}
MIPS & $-9.61 \pm 0.082$  & $-0.366 \pm 0.072$  \\
SED & $-9.95 \pm 0.099$  & $-0.755 \pm 0.12$  \\
OII & $-9.82 \pm 0.078$  & $-0.530 \pm 0.073$ \\
\cutinhead{Cluster removed (Figure~\ref{ssfr_density_out_all})}
MIPS & $-9.73 \pm 0.096$  & $-0.246 \pm 0.10$  \\
SED & $-9.98 \pm 0.12$  & $-0.653 \pm 0.12$  \\
OII & $-9.82 \pm 0.094$  & $-0.534 \pm 0.098$  \\
\cutinhead{$UVJ$-selected star forming galaxies (Figure~\ref{ssfr_density_sfs_all})}
MIPS & $-9.46 \pm 0.086$  & $-0.0787 \pm 0.090$  \\
SED & $-9.61 \pm 0.11$  & $-0.472 \pm 0.11$  \\
OII & $-9.42 \pm 0.099$  & $-0.438 \pm 0.10$ 
\enddata
\end{deluxetable}

\begin{figure}
\epsscale{1.2}
\plotone{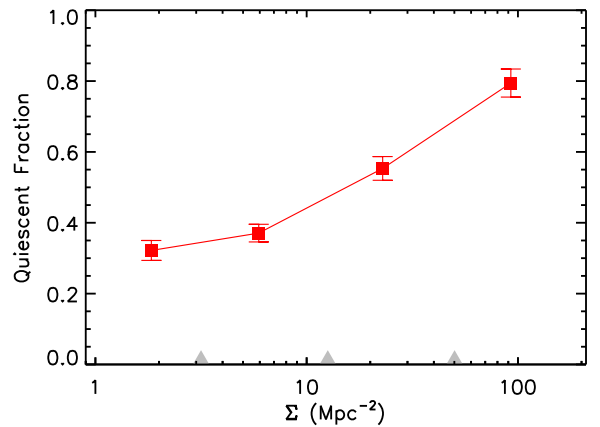}
\caption{Fraction of $UVJ$-selected quiescent galaxies versus density for galaxies with mass \mcutrangelog\ at \zwindowwide.  The fraction of galaxies in the quiescent clump more than doubles from $32\%$ at the lowest densities to $79\%$ in the cores of groups and a cluster. \label{frac_qc_density}}
\end{figure}

Here, we investigate how a changing proportion of quiescent and SFGs with density contribute towards the declining SSFR-density relation found in \S~\ref{sfr_sec_sfrdensity}.  Figure~\ref{frac_qc_density} shows the fraction of $UVJ$-selected quiescent galaxies at \zwindowwide\ as a function of local density using the mass-limited sample.  As one might expect from the increasing red galaxy fraction \citep[][]{patel2009} for this sample, galaxies at higher densities are more likely to be in the quiescent clump.  The fraction of galaxies in the quiescent clump increases from $32 \pm 3\%$ at low density to $79 \pm 4\%$ at high density.  The quiescent fraction remains fairly constant across the two lowest density bins, which are representative of the field.  The fraction begins to increase substantially between the middle two density bins.  Figure~\ref{densitymap} shows that these two density bins represent the transition between what are likely filaments, small groups, and the outskirts of larger groups (green) to the centers of large groups and the outskirts of clusters (orange).  Several other works at these redshifts also find evidence for a lower {\em fraction} of actively star forming systems in higher density environments, based on H$\alpha$ emission \citep{sobral2011}, [OII] emission \citep{poggianti2008} or mid-IR emission \citep{finn2010} \citep[or both emission lines and mid-IR flux at somewhat lower redshifts, e.g.,][]{tran2009,balogh2009}.

Finally, we note that when using a similar mass cut as in \citet{vanderwel2007b}, the quiescent fraction follows the early-type galaxy fraction vs. density from that work remarkably well.

\subsection{The Role of Star-Forming Galaxies in the SSFR-Density Relation} \label{sec_ssfr_sfg}

\begin{deluxetable}{lcc}
\tablewidth{0pc}
\tablecolumns{3}
\tablecaption{Parameters for best-fit $\log-\log$ lines for SSFR-density relations in Figure~\ref{ssfr_density_all_massbin} \label{table_ssfr_density_all_massbin}}
\tablehead{
\colhead{Mass range} & \colhead{Intercept} & \colhead{Slope} \\
\colhead{} & \colhead{($\log$~SSFR)} & \colhead{($\Delta(\log$~SSFR)/$\Delta(\log \Sigma$))}
}
\startdata
\cutinhead{SED}
$10.25 < \log M/M_{\odot} < 10.50$ & $-9.55 \pm 0.19$  & $-0.810 \pm 0.19$  \\
$10.50 < \log M/M_{\odot} < 10.80$ & $-10.1 \pm 0.20$  & $-0.721 \pm 0.16$  \\
$10.80 < \log M/M_{\odot}$ & $-10.2 \pm 0.12$  & $-0.674 \pm 0.13$  \\
\cutinhead{OII}
$10.25 < \log M/M_{\odot} < 10.50$ & $-9.36 \pm 0.14$  & $-0.660 \pm 0.12$  \\
$10.50 < \log M/M_{\odot} < 10.80$ & $-9.87 \pm 0.14$  & $-0.499 \pm 0.13$  \\
$10.80 < \log M/M_{\odot}$ & $-10.1 \pm 0.11$  & $-0.375 \pm 0.11$  
\enddata
\end{deluxetable}

The analysis in \S~\ref{sfr_sec_sfrdensity} for all galaxies above the mass limit of \mcutrangelog, establishes that the median SSFR and SFR decline in higher density environments at \zwindowwide, confirming the declining trend found in \citet[][]{patel2009b}.  In the previous section, we found that the changing proportion of quiescent and SFGs contributes towards the declining SF activity at higher densities.  Here, we investigate whether a change in the SFRs of $UVJ$-selected SFGs contributes towards the declining SSFR-density relation.

Utilizing the $UVJ$ diagram (Figure~\ref{sfr_fig_UVJ_density}), we now select SFGs.  Figure~\ref{ssfr_density_sfs_all} shows the SSFR-density relation, derived from all three SFR indicators, for SFGs above the mass limit at \zwindowwide\ with $K_s$ imaging.  The SSFRs of SFGs decline at higher densities for the SED and [OII] derived SFRs by factors of $\sim 5-6$, while the MIPS derived SSFRs decline only by a factor of $\sim 1.3$ across the full range of densities.  Straight line fits to the MIPS, SED, and [OII] log(SSFR)-log(density) relations indicate non-zero, negative slopes at significances of $\sim 0.9$, $4$, and $4\sigma$ (see Table~\ref{table_ssfr_density} for best fit parameters).  We note that if galaxies at \zwindow\ are excluded (i.e., those near the cluster), the MIPS SSFRs for SFGs decline by a factor of $\sim 1.5$ from the lowest to highest densities but the negative slope from a straight line fit ($\Delta(\log$~SSFR)/$\Delta(\log \Sigma) \approx -0.10$) is only significant at the $\sim 1.0 \sigma$ level.  The decline in the SSFR-density relation, especially for the SED and [OII] derived relations, is therefore driven in part by declining SFRs of SFGs, in addition to the increasing fraction of quiescent galaxies at higher densities. We note that at a given density the median SSFRs for quiescent galaxies are typically more than an order of magnitude lower than the value for SFGs for a particular SFR indicator, and generally lack any significant dependence on density.

\subsection{A Parallel Decline with Density in Star-Formation and Dust Content} \label{sec_av_density}

\begin{figure}
\epsscale{1.2}
\plotone{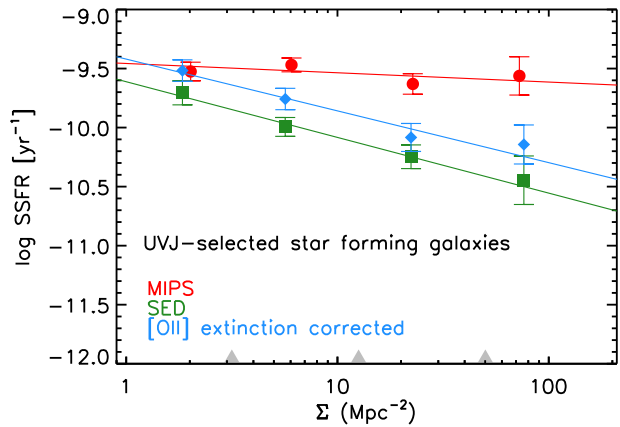}
\caption{Median SSFR versus local density for SFGs with mass \mcutrangelog\ at \zwindowwide.  All three SFR indicators find the SSFRs of SFGs to decrease at higher densities.  The decline in the SSFR-density relation for all galaxies (Figure~\ref{ssfr_density_all}) is therefore caused by declining SFRs for SFGs as well as a change in the proportion of quiescent and SFGs (Figure~\ref{frac_qc_density}).  Note that the decline in SSFR with density for the MIPS derived SFRs is not as strong as the other two SFR indicators. \label{ssfr_density_sfs_all}}
\end{figure}

\begin{figure}
\epsscale{1.2}
\plotone{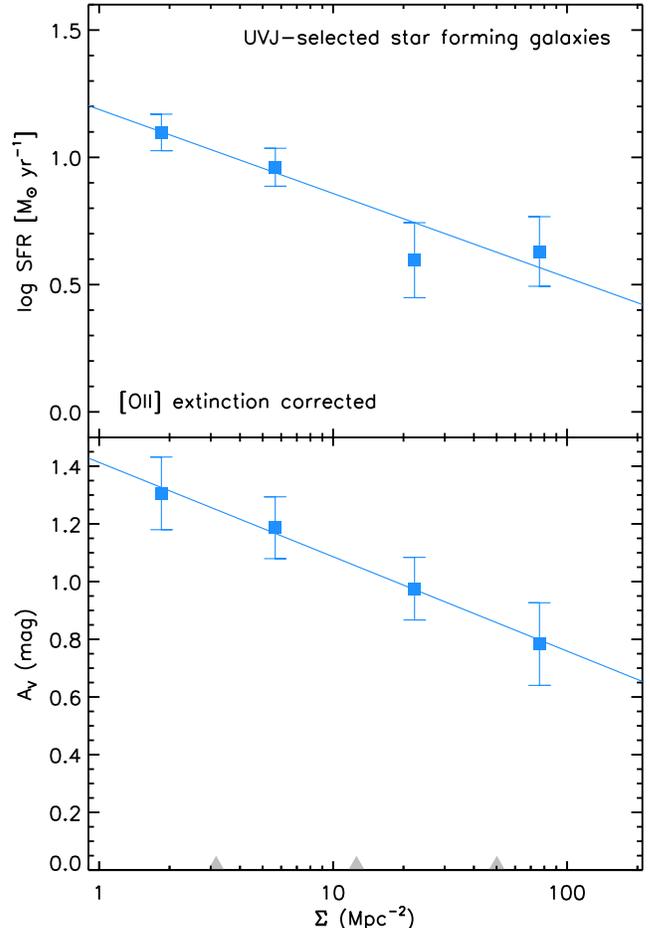}
\caption{(a) Median extinction corrected [OII] SFR versus local density for SFGs with mass \mcutrangelog\ at \zwindowwide. (b) Median \av\ from SED fit versus local density for the same sample of SFGs.  Both the attenuating dust content and SFRs of SFGs decline at higher densities, as one would expect if the cold gas supply is diminished in such environments. \label{av_density_sed}}
\end{figure}

SF and dust production are connected since most dust is injected into the ISM through AGB stars or supernovae, both the by-products of recent episodes of SF \citep[see, e.g.,][]{mathis1990}.  Subsequent episodes of SF are also dependent on the presence of dust in allowing gas to cool and form stars.  With the SFRs of SFGs decreasing with increasing density, we should expect to see less dust associated with SFGs at higher densities.  

Figure~\ref{av_density_sed}$a$ shows the extinction corrected [OII] SFR-density relation for the same sample of SFGs used in Figure~\ref{ssfr_density_sfs_all}.  A straight line fit to the log(SFR)-log(density) relation gives a slope of $\Delta(\log$~SFR)/$\Delta(\log \Sigma) \approx -0.33$ and is significant at $\sim 3.6\sigma$.  The decline in SFR from low to high density is a factor of $\sim 3$, somewhat smaller than the factor of $\sim 5$ decline in SSFR (Figure~\ref{ssfr_density_sfs_all}).  However, within the uncertainties on the best-fit slopes, these numbers are compatible.  Note also that the median mass is slightly higher at higher densities (see blue numbers in Figure~\ref{sfr_fig_UVJ_density}), leading to a steeper slope for the SSFR-density relation.  The [OII] derived SFRs at high densities could also be overestimated due to AGN activity \citep[see e.g.,][]{lemaux2010,kocevski2010b}, further strengthening the declining SSFR-density and SFR-density trends found here.

Meanwhile, Figure~\ref{av_density_sed}$b$ shows the median value of the best-fitting \av\ from the SED fits versus density for the same SFGs.  The lowest density environments at \zwindowwide\ are associated with the highest dust content, with the median \av\ decreasing by $\sim 0.5$~mag from low to high density.  A straight line fit to the \av-log(density) relation indicates a slope of $\Delta (A_V)/\Delta (\log \Sigma) \approx -0.32$~mag~dex$^{-1}$ and is significant at the $\sim 3.0\sigma$ level.  We note that the median \av\ of quiescent galaxies remains constant across all densities, suggesting that the relative differences in \av\ for SFGs are credible.  The parallel decline in SFR and \av\ with density for SFGs suggests that the cessation of SF in higher density environments is linked with the level of attenuating dust.  The lower dust content at higher density could be a consequence of a physical mechanism removing the dust, or a lack of replenishment due to lower levels of SF.

\section{Discussion} \label{sfr_sec_discussion}

\subsection{No Reversal in the Mass-limited SSFR-Density and SFR-Density Relations at $z \lesssim 1$}

At \zwindowwide, we find that at a fixed stellar mass, the median SSFR and SFR declines from low to high density environments for three different SFR indicators, as also found with MIPS based SFRs in \citet{patel2009b}.  Another new key result is that we find this declining SFR-density relation persists even when excluding galaxies in the vicinity of the cluster (Figure~\ref{ssfr_density_out_all}), indicating that the declining SFR trend with density applies more generally to all environments.  As noted in \citet{patel2009b}, our result differs from that of \citet{elbaz2007} and \citet{cooper2008}, both of which find a reversal in the SFR-density relation at $z \sim 1$ for luminosity-limited samples.  The reversal found by these other works does not extend to the highest densities probed by their surveys.  Instead, their SFR-density relations turnover and decrease at intermediate densities and above.  We note however that we do not find any intermediate density regime in which the median SFR is significantly elevated relative to the field.  Our results indicate that for galaxies at \zwindowwide\ with stellar masses \mcutrange, SF activity in groups and clusters is lower than in the field, similar to what is found at lower redshifts \citep{wolf2009} and at $z \sim 0$ \citep{kauffmann2004}.  Other work at $z \sim 1$ also find a declining SFR or SSFR-density relation \citep{scoville2007c}.

The size of the density bins used in this study are fairly broad ($\sim 0.5$~dex).  Is it possible that at the intermediate densities where a reversal has been suggested, our density bins lack the resolution required to sample any elevated SF activity?  The reversal in \citet{cooper2008} occurs at overdensities relative to the median density of $\sim 0.06$ to $\sim 5$.  This translates into a density range that roughly covers our three lowest density bins.  The reversal in \citet{elbaz2007} extends to densities that are a factor of $\sim 5-6$ higher than their lowest densities.  This would be encompassed by our two lowest density bins.  Any reversal would therefore be observable in our dataset even with our broad density bins, but is not.

Our results show that the epoch at which a reversal in the SFR-density relation takes place, if any, is still under debate.  Above a mass limit of \mcutrangelog, we find that the SFR-density relation does not reverse at $z \lesssim 1$.  When studying galaxies at $z \sim 0.85$ in a fixed mass range, \citet{cooper2010b} find a larger proportion of red galaxies at high densities compared to low densities.  Assuming these red colors reflect older stellar populations rather than dusty SF, their findings lend support to the declining SFR-density relation found in \citet{patel2009b} and in this work.  The debate also extends to higher redshifts as studies of environment have begun to expand beyond $z>1$.  \citet{tran2010b} find evidence for a high fraction of SFGs associated with a cluster at $z \sim 1.6$.  Also, \citet{hayashi2010} find [OII] derived SFRs for galaxies in a cluster at $z \sim 1.5$ that are similar to those of galaxies in the outskirts.  In contrast, \citet{gobat2010} find a well formed X-ray emitting cluster with a large red elliptical galaxy population at $z \sim 2$.  Also at $z \sim 2$, \citet{tanaka2010b} find the SFRs of galaxies in a proto-cluster to be lower than that of galaxies in the field.  Meanwhile, wide-field clustering studies at $2<z<3$ find red galaxies to be strongly clustered, suggesting that a color-density relation was in place at those early times with red galaxies dominating more massive halos \citep{quadri2007,quadri2008}.  These works point out however that one key unresolved question is the source of the red colors: dusty SF or old stellar populations.  As a followup, Quadri et~al. (2011, in preparation) find the fraction of quiescent galaxies, selected with a $UVJ$ diagram similar to ours, to increase with density out to $z \sim 2$.

\subsection{The Origin of the SFR-Density Relation at \zwindowwide}

One of the key questions we set out to answer was how the SFRs of individual galaxies were changing in order to produce the declining SSFR-Density relation (Figure~\ref{ssfr_density_all}).  Either we are seeing (1) a changing mix of quiescent and SFGs with increasing density, (2) a decline in the SFRs of SFGs with density, or (3) a combination of the two effects.

From Figure~\ref{frac_qc_density}, we infer that the fraction of SFGs declines from $\sim 68\%$ at the lowest densities to $21\%$ at the highest densities, a factor of $\sim 3$ decrease.  If the SFRs of SFGs were to remain constant across various environments, and the SFRs of quiescent galaxies were negligible, then the change in the relative proportion of quiescent and SFGs with environment would contribute only a factor of $\sim 3$ towards the decline in the SSFR-density relation.  This amount is less than the order of magnitude decline found in Figure~\ref{ssfr_density_all} for the SED and [OII] based SFRs, but close to the factor of $\sim 4$ decline found for the MIPS based SFRs.  The change in the fraction of quiescent and SFGs with density therefore cannot be solely responsible for driving the declining SED and [OII] SSFR-density relations.  Indeed, this is what we find in Figure~\ref{ssfr_density_sfs_all}, with the SFRs of SFGs declining by factors of $\sim 5-6$.  The MIPS based SFRs of SFGs also decline, but by a much smaller amount.  Thus, both (1) and (2) above contribute towards the declining SSFR-density relation and potentially reflect a diversity of mechanisms that are responsible for quenching SF on different timescales \citep[see, e.g.,][]{treu2003,moran2007c}.

Other environmental studies at somewhat lower redshifts also find the SFRs of SFGs to decline at higher densities.  For example, using a different color-color selection from our own (in addition to other selection criteria) to identify SFGs, \citet{wolf2009} found the SSFRs of SFGs to decline at higher densities at $z \sim 0.17$.  Meanwhile, \citet{poggianti2008} argue that SF activity in SFGs at $0.4<z<0.8$ is independent of local density based on [OII] equivalent widths.  However, a straight line fit to the $\log$(SSFR)-$\log$(density) relation in Figure~7 of \citet{poggianti2008} indicates a negative slope of $\Delta(\log$~SSFR)/$\Delta(\log$~density$) \approx -0.5$ at a significance of $\sim 4\sigma$.  We note that our definition for SFGs differs from that of \citet{poggianti2008}.  Also, while the luminosity-limited sample of \citet{poggianti2008} is limited to regions in the vicinity of clusters, our survey samples a broader range of environments.

The decline in SFRs with increasing density for SFGs seen in Figure~\ref{ssfr_density_sfs_all} could suggest a gradual shut down in the cold gas supply that fuels SF at \zwindowwide.  This is further supported by the decreasing value for \av\ with increasing density.  As dust and cold gas are tied to the SF process, the declining \av\ at higher densities suggests less fuel for SF: recent observations of HI deficient spirals in the Virgo cluster with truncated dust disks \citep{cortese2010c,delooze2010} provide strong support for this view.  Other recent works at low redshift also point to the slow removal of gas in order to reproduce various observations \citep[e.g.,][]{weinmann2009,weinmann2010,vanderwel2010}.

Finally, we note that our selection of SFGs in this work utilizes the color-color space in which they are predicted and observed to occupy.  Other works that require the detection of emission lines or mid-IR flux find that such galaxies have relatively little variation in SFR at a fixed mass (intrinsic $1-\sigma$ scatter of $\sim 0.3-0.5$~dex) \citep[see, e.g.,][]{brinchmann2004,noeske2007,rodighiero2010b}.  Meanwhile, the range in SFRs (SED and [OII]) for our $UVJ$-selected SFGs across our three lowest density bins (i.e., environments typically probed by the field surveys above) is roughly $\sim 0.6$~dex.  Thus, our selection of SFGs and interpretation of their properties in this work may differ from those of others.

\subsection{SFRs of SFGs from MIPS vs. Other SFR Indicators}

The SSFR-density slope for SFGs in Figure~\ref{ssfr_density_sfs_all} for MIPS based SFRs is much shallower compared to the SED and [OII] based SFRs.  Several recent results could explain these findings.  For example, while the overall SF activity is clearly diminished in higher density regions, the remaining SF that takes place may occur in a more obscured state.  \citet{wolf2009} find that the few SFGs at high densities at $z \sim 0.17$ that remain are predominantly spirals of the reddened variety \citep[see also,][]{gallazzi2009,haines2010}.  \citet{bekki2010e} propose that such red, spiral SFGs in higher density environments result from ``outside-in truncation'', with the remaining SF taking place in the inner regions of the galaxy where higher gas densities and metallicities lead to heavy obscuration.  Another explanation could be that increased AGN activity in high density regions at these redshifts \citep{martini2009} could be responsible for the elevated mid-IR flux.  Our LDP spectra lack the resolution necessary to identify Type 2 AGN from emission line diagnostics and therefore such objects likely contaminate our sample.  We note however that less than $<1\%$ of our sample has an X-ray counterpart from the ChaMP point source catalog \citep{kim2007b}, and ignoring these objects does not impact our results.  In addition, \citet{fu2010} used {\em Spitzer} IRS spectra to show that objects at $z \sim 0.7$ with MIPS 24~\micron\ fluxes less than $\sim 1$~mJy, as is the case with our entire sample, are dominated by SF rather than AGN activity.  Finally, the shallow decline of SSFR with density for SFGs in Figure~\ref{ssfr_density_sfs_all} from the MIPS based SFRs may reflect the potentially longer timescales probed by that SFR indicator \citep[see, e.g.,][]{salim2009,kelson2010}.  High density regions accrete galaxies from lower densities where the typical SFRs are much higher.  If a SFR indicator traces SF over long timescales, the median SFR for a sample of galaxies at high densities will be higher given that the newly accreted galaxies from low density will still register high SFRs.  Lending further support to this explanation is the finding that MIPS 24~\micron\ detected sources at high densities at these redshifts exhibit strong Balmer absorption \citep{kocevski2010}.  This would indicate a substantial population of A-stars of age $\sim 1$~Gyr old, a timescale in accord with the findings of \citet{salim2009} and \citet{kelson2010}.

\section{Summary} \label{sfr_sec_summary}

We have carried out a wide-field spectroscopic survey of galaxies at \zwindowwide\ that span a range of environments, including the low density field, groups, and the $z \sim 0.83$ cluster \rxj.  We selected galaxies with $z_{\rm AB}<23.3$~mag for spectroscopy with IMACS on Magellan with a low-dispersion prism (LDP).  These LDP spectra yielded redshifts with a remarkable precision of $\sigma_z/(1+z) \sim 1\%$.  

We used a mass-limited sample above \mcutrangelog\ to study the SFRs of galaxies as a function of the local galaxy density.  Above the mass limit, the distribution of stellar masses in different density bins in our sample is very similar (Figure~\ref{frac_qc_density}).  Thus the environmental trends reported in this paper are not significantly affected by any residual correlation with stellar mass.

Given the systematics associated with any single SFR indicator, we used three different SFR indicators in drawing conclusions about the environments at \zwindowwide\ where SF is quenched.  We computed SFRs from: (1) rest-frame 12-15~\micron\ luminosities measured with {\em Spitzer} MIPS 24~\micron\ imaging, (2) SED fitting, and (3) extinction corrected \oii\ luminosities from the LDP spectra.

To distinguish SFGs from quiescent galaxies we used a $U-V$ vs. $V-J$ diagram (referred to here as the $UVJ$ diagram).  This allowed us to determine whether the SFR-density relation at \zwindowwide\ was driven by (1) a change in the proportion of quiescent and SFGs or (2) a change in the SFRs of SFGs in different environments.  The $UVJ$ diagram is especially useful in isolating dusty, SFGs on the red-sequence, which would otherwise be classified as quiescent based on a single optical color (Figures~\ref{bicolor} and \ref{uvj_hst}).  

Our main results are the following:

\begin{enumerate}
\item With three different SFR indicators, we find that the median SSFR-density relation at \zwindowwide\ declines across the entire range of densities probed: from the low-density field, to the cores of groups and a cluster (Figure~\ref{ssfr_density_all}).  The declining SSFR-density relation holds even when removing galaxies near the cluster (Figure~\ref{ssfr_density_out_all}).  The declining SSFR-density relation is therefore general to all environments at these redshifts and not driven by galaxies located near the rich cluster in our sample.  

\item For fixed mass bins, the SSFR-density shows a clear decline with increasing density (Figure~\ref{ssfr_density_all_massbin}), confirming that the local environment is indeed a factor in impacting the SFRs of galaxies when controlling for mass.  This result also implies that at a fixed mass the SFR-density relation at these redshifts follows a similar decline with increasing density.

\item The fraction of quiescent galaxies increases by a factor of $\sim 2.5$, from $\sim 32\%$ to $\sim 79\%$, between the lowest and highest densities (Figure~\ref{frac_qc_density}).  We note that these quiescent fractions are similar to the morphological early type fractions seen in \citet{vanderwel2007b} when adjusting for the mass selection.  We also find that the SSFRs of SFGs decline with density (Figure~\ref{ssfr_density_sfs_all}).  The declining SFR-density and SSFR-density relation at \zwindowwide\ can therefore be attributed to a combination of (1) a changing mix of quiescent and SFGs, and (2) a decline in the SFRs of SFGs with increasing density.

\item In parallel with the declining SFRs of SFGs, the median \av\ of SFGs also decline at higher densities (Figure~\ref{av_density_sed}$b$).  This decline in \av\ signifies a lack of cold gas and therefore the quenching of SF in higher density environments.

\end{enumerate}

In future work, we plan to use HST ACS and WFC3 imaging to determine how SF activity is distributed within SFGs in different environments.  Such an analysis will lead to insights on whether SF is being turned off across the entire galaxy (e.g., via starvation) or only in the outer regions (e.g. gas stripping).  The answer has implications for the mechanisms that are responsible for regulating SF in galaxies residing in different environments at half the age of the universe.

\acknowledgements
We thank Sandra Faber and David Koo for helpful discussions.  We also thank Stijn Wuyts for providing a table for converting MIPS 24~\micron\ fluxes into total infrared luminosities.  We also wish to acknowledge those who have contributed to the construction and deployment of IMACS as well as Scott Burles for developing the low-dispersion prism, and the PRIMUS collaboration for allowing us to investigate galaxies with their hardware.  This research was supported by NASA grant NAG5-7697 and Spitzer grant JPL 1277397.

\begin{appendix}

\section{Stellar Mass Limit of the Sample} \label{app_masslimit}
\begin{figure}
\epsscale{0.8}
\plotone{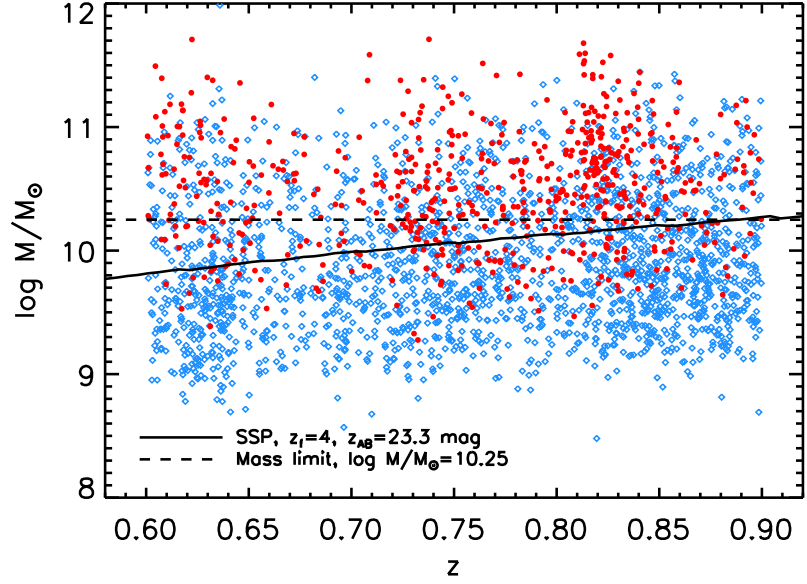}
\caption{Stellar mass vs. redshift for galaxies at \zwindowwide\ with $z_{\rm AB}<23.3$~mag and with $K_s$ imaging.  $UVJ$-selected quiescent galaxies are indicated by the solid red circles, while SFGs are shown as open blue diamonds.  The solid black line indicates the stellar mass of a galaxy that has undergone an instantaneous burst at $z_f=4$ and is observed to have $z_{\rm AB}=23.3$~mag at all redshifts.  This model indicates that our sample can be made representative for young SFGs {\em and} old quiescent galaxies above a limiting stellar mass of \mcutrangelog\ (dashed black line).  Note that the mass completeness limit for SFGs (blue points) is much lower. \label{masslimit}}
\end{figure}

We primarily use a stellar mass limited sample to conduct our study.  Figure~\ref{masslimit} illustrates how we determine our mass limit.  The figure shows the stellar masses of galaxies with $z_{\rm AB}<23.3$~mag as a function of redshift for galaxies at \zwindowwide\ and with $K_s$ imaging.  Using the $UVJ$ diagram (see Figure~\ref{bicolor}), we identify quiescent galaxies and denote them as red circles in the figure.  SFGs are shown as open blue diamonds.  Note how SFGs extend to low masses where low mass quiescent counterparts are lacking.  In order to study a representative stellar mass limited sample of quiescent and SFGs the mass limit will then be determined by the redder quiescent galaxies.  The solid line shows the stellar mass for an old stellar population formed at $z_f = 4$ and observed to have a magnitude of $z_{\rm AB}=23.3$~mag at all redshifts.  Note that a similar galaxy with a fainter magnitude would fall below this line.  For a quiescent galaxy in our sample at $z_{\rm AB}=23.3$~mag (i.e., our survey magnitude selection limit) the limiting stellar mass at $z=0.9$ is \mcutrangelog\ (dashed line).  This value represents the limiting stellar mass for galaxies in the redshift interval studied in this paper, \zwindowwide.

\section{Spectroscopic Completeness}
\begin{figure}
\epsscale{1}
\plotone{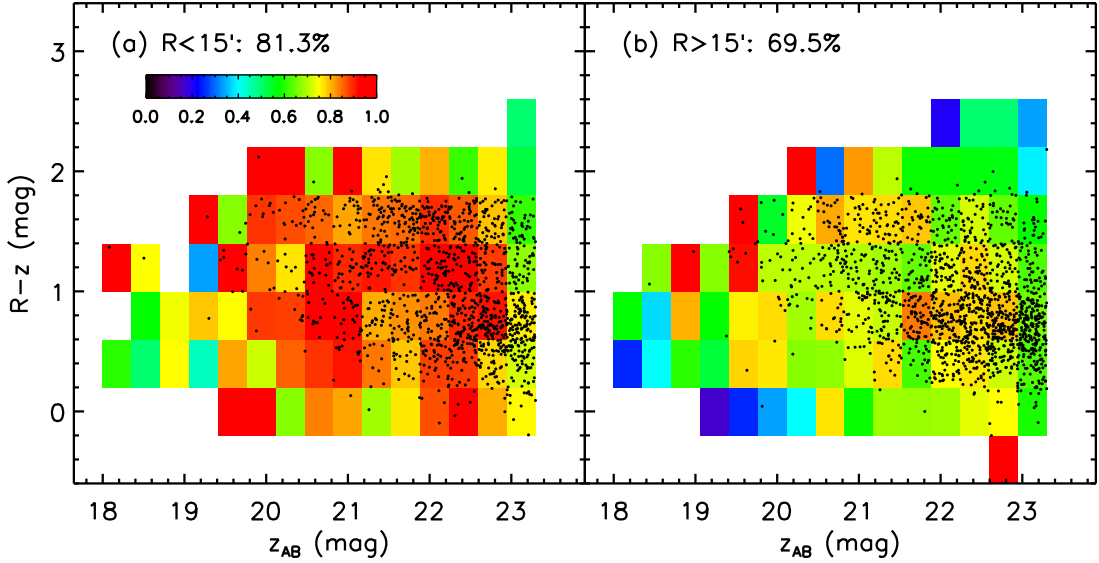}
\caption{Spectroscopic completeness as a function of observed $R-z$ color and $z$-band magnitude for ($a$) the inner $R \lesssim 15\arcmin$ and ($b$) the outer $R \gtrsim 15\arcmin$ regions of the $z$-band imaging (see text for precise boundaries of each region).  The overall completeness for each region is indicated in the top left of each panel.  A color bar corresponding to the completeness fraction is shown in the top left.  We note that the $R-z$ magnitudes represent Vega$-$AB mag as the raw data were zeropointed using the system that was native to each filter ($R_{\rm AB}=R_{\rm Vega}+0.185$~mag).  Each panel also shows the $R-z$ color and $z$-band magnitudes of galaxies in the given region at \zwindowwide, giving a sense for where most of our sample lies in these completeness maps.  \label{fig_compmap}}
\end{figure}

In making various measurements in this paper, we have taken care to account for variations in the spectroscopic completeness of our survey.  The completeness here is defined as the ratio of the number of galaxies with a high quality LDP redshift to the number of galaxies in the $z$-band selection catalog.  The reciprocal of the completeness fraction is used as a weight when computing quantities such as the local galaxy density, median SSFRs, etc.  Figure~\ref{fig_compmap} shows the spectroscopic completeness as a function of the observed $R-z$ color and $z$-band magnitude.  Panel ($a$) shows the completeness within an ellipse centered on RA$_0=28.1814$~deg and Dec$_0=-13.9998$~deg, with axis lengths $a_{\rm RA}=0.174$~deg and $b_{\rm Dec}=0.132$~deg.  Panel ($b$) shows the completeness in the outer regions with the same RA$_0$ and Dec$_0$ but with $a_{\rm RA}=0.29$~deg and $b_{\rm Dec}=0.22$~deg.  In general, the inner region has a slightly higher level of completeness.  Overplotted on each completeness map are the colors and magnitudes of objects within the given region and with redshifts in the range \zwindowwide.  The bulk of our sample lies in regions of completeness space that are relatively smooth.

\end{appendix}


\clearpage


\end{document}